\documentclass{article}
\usepackage{jheppub} 
\usepackage{amsmath,amssymb} 
\usepackage{rotating}

\title{Gravitational instantons from closed superstring field theory}

\author[a,1,2]{Ivo Sachs}
\author[a,b,1,2]{and Xianghang Zhang}
\affiliation[a]{Arnold{-}Sommerfeld{-}Center for Theoretical Physics, Ludwig{-}Maximilians{-}Universit\"at M\"unchen, Theresienstr. 37, D{-}80333 M\"unchen, Germany}
\affiliation[b]{Graduate School of Mathematics, Nagoya University, Furocho 1, Nagoya 464{-}8602, Japan}
\note[1]{These authors contributed equally to this work.}
\emailAdd{ivo.sachs@physik.lmu.de}
\note[2]{ORCID identifiers: Ivo Sachs, 0000-0003-2526-8655; Xianghang Zhang, 0009-0006-0951-2119.}
\emailAdd{zhang.xianghang.g5@math.nagoya-u.ac.jp}
\abstract{We test exact marginality of the deformation describing the resolution of a $\mathbb{Z}_{2}$ orbifold by analyzing the closed superstring equations of motion to third order in the size, including $\alpha'$ corrections. We find that the third order correction is unobstructed for all deformation moduli. We are also able to reproduce the Eguchi-Hanson gravitational instanton up to the second order in the field theory limit with a suitable choice of moduli.}

\begin{document}

\maketitle
\flushbottom

\section{Introduction}
It is established that string theory provides a way to resolve the orbifolds by ``blowing up'' the singular points to get a more general background~\cite{Candelas1985,Aspinwall1994}.
For example, the K3 manifold, as a specific class of Calabi-Yau manifolds, can be obtained from the quotient space of a complex torus, yielding promising phenomenology.
However, explicit calculations on a resolved background are complicated since metrics on Calabi-Yau manifolds are not well understood.
Therefore, the orbifold model is an ideal starting point for approaching a generic string background via exactly marginal deformation of the worldsheet conformal field theory (CFT).

The linear deformations corresponding to resolving orbifold singularities can be identified with marginal operators built from worldsheet twist fields.
Whether this marginal deformation can be lifted to a full classical background is the problem of exact marginality.
A similar analysis for D-brane configurations and boundary twist fields has appeared in~\cite{Billó2003}, where it was argued that it was essentially an off-shell problem.
A more suitable framework to deal with off-shell string theory is string field theory~\cite{Sen2015,deLacroix2017} (SFT).
The D-brane instanton background has been analyzed in open superstring theory~\cite{Mattiello2019,Maccaferri:2019ogq,Vošmera2019}.
In this paper, we use closed superstring field theory developed in \cite{Erler2014} to approach the resolution of orbifold singularities in string theory, and obtain the connection to classical gravitational instantons in the field theory limit.
Our main results are as follows.
 First, we show that the marginal deformation is unobstructed at the second order in Type II superstring field theory, bypassing the tachyonic and obstruction issues inherent in the bosonic case.
 Extracting the spacetime fields in the field theory limit $\alpha' k^{2} \ll 1$, we explicitly reproduce the Eguchi-Hanson metric $W_{ij}^{(2)}$ with a particular choice of moduli.
 Furthermore, we prove that the induced second-order geometry is inherently hyperk\"ahler by verifying the self-duality of the Weyl tensor $C = C^{+}$.
 Finally, we analyze the equations of motion to the third order and demonstrate that the cohomological obstruction $\mathcal{O}^{(3)}$ vanishes identically for all deformation moduli, which contrasts with the open string sector where generalized ADHM constraints are typically required.

The paper is organized as follows.
In section~\ref{sec:orb}, we review the Type IIB superstring spectrum on the $\mathbb{C}^{2}/\mathbb{Z}_{2}$ orbifold and identify the relevant vertex operators corresponding to the marginal deformation.
Section~\ref{sec:sft} outlines the classical action and the $L_{\infty}$ algebraic structure of the NS-NS closed superstring field theory.
In section~\ref{sec:res}, we perform the perturbative analysis of the orbifold resolution, calculating the second and third-order obstructions and deriving the induced second order spacetime metric.
We conclude in section~\ref{sec:con} with a discussion.
In appendix~\ref{app:obs} we show that the deformation is obstructed already at the second order in the bosonic theory, and hence we consider only Type II theory in the main text.
Appendix~\ref{app:geo} is a brief review of hyperk\"ahler geometry and asymptotically locally Euclidean (ALE) gravitational instantons.
We also show that the second order metric correction is hyperk\"ahler.
Appendix~\ref{app:sft} is devoted to the explicit definition of the closed string field theory vertices.
The explicit $U(1)$-charge decomposition establishing the vanishing of the third-order obstruction is presented in appendix~\ref{app:third-order}.

\section{Superstring theory on an orbifold}\label{sec:orb}
In this section, we briefly review the spectrum of Type IIB superstring theory on the orbifold with particular focus on the role of twist fields.\footnote{Apart from the well known problem of the tachyonic vacuum in the bosonic theory, another reason for considering only superstrings is shown in appendix~\ref{app:obs}, where we see the deformation in bosonic theory is obstructed.}
Concretely, we consider a  $\mathbb{C}^{2}/\mathbb{Z}_{2}$ conifold and will identify the marginal vertex operator corresponding to the resolution of the orbifold singularity.
The four real coordinates $(x^{1}, x^{2}, x^{3}, x^{4})$ on $\mathbb{C}^{2}$ are identified with $(-x^{1}, -x^{2}, -x^{3}, -x^{4})$ via a $\mathbb{Z}_{2}$ group action.
In the worldsheet conformal field theory, the orbifold structure is obtained by considering twisted sectors in addition to the untwisted sector~\cite{Dixon1985,Dixon1987}.
The twisted sectors are built on the vacuum by action of a twist field $\sigma$ (and $S^\alpha$), thereby implementing the $\mathbb{Z}_{2}$ twist.
We focus on the NS-NS sector which contains the geometric moduli of interest.
\subsection{The untwisted sector}
The untwisted sector $\mathcal{H}_{0}$ consists of states that are invariant under the $\mathbb{Z}_{2}$ group action.
These are states that are present in the theory before orbifold projection.
In the NS-NS sector, the ground state energy is $E_{0} = -1/2$ and the GSO projection keeps the massless state.
We denote the four bosonic fields along uncompactified Euclidean directions as $X^{i}(z, \bar{z}),\,i \in \{1, 2, 3, 4\}$.
The untwisted moduli come from the $(\mathrm{NS}+, \mathrm{NS}+)$ sector of momentum $k$,
\begin{equation}
W_{ij}(k) \psi^{i}_{-\frac{1}{2}} e^{ikX} \bar{\psi}^{j}_{-\frac{1}{2}} e^{ik\bar{X}}|0\rangle_{\mathrm{NS}} |\bar{0}\rangle_{\mathrm{NS}}.
\end{equation}
The spacetime field $W_{ij}$ contains the graviton, the dilaton and the antisymmetric Kalb-Ramond field.

\subsection{The twisted sector}

In the twisted NS-NS sector $\mathcal{H}_{\theta}$, the closed string boundary conditions are twisted,
\begin{equation}
X^{i}(\sigma + 2\pi) = -X^{i}(\sigma),
\end{equation}
which shifts the mode expansion.
The fermions $\psi^{i}$ are integer moded and we have ground state energy $E_{0} = 0$, leading to four zero modes for each chirality.
This gives rise to four twisted moduli
\begin{equation}
\bar{w}_{\alpha} w_{\beta} \bar{S}^{\alpha} S^{\beta} \bar{\Delta} \Delta |0\rangle_{\mathrm{NS}} |\tilde{0}\rangle_{\mathrm{NS}},
\end{equation}
where $\alpha, \beta \in \left\{1, 2\right\}$, $\Delta = \sigma^{1}\sigma^{2}\sigma^{3}\sigma^{4}$, and $\bar{\Delta} = \bar{\sigma^{1}}\bar{\sigma^{2}}\bar{\sigma^{3}}\bar{\sigma^{4}}$.
These are the marginal deformations that give a basis of the tangent space at the origin of the moduli space of the (uncompactified) Kummer surface resulting from the deformations of the orbifold.
Taking into account the ghost sector, the corresponding marginal vertex operator in the $(-1, -1)$ picture is
\begin{equation}\label{eqn:md}
\mathcal{V} = \bar{c}c\bar{V}Ve^{-\phi}e^{-\bar{\phi}}
\end{equation}
with
\begin{equation}
V = w_{\beta} S^{\beta} \Delta, \quad \bar{V} = \bar{w}_{\beta} \bar{S}^{\beta} \bar{\Delta}.
\end{equation}
This is to be identified with the first order solution in our following string field theory treatment.

\section{Classical closed superstring field theory in the NS-NS sector}\label{sec:sft}
The string field theory action we use is described in~\cite{Erler2014} but we also adopt notations to reconcile with more recent literature~\cite{deLacroix2017, Sen2024}.
We will only make use of the action in the NS-NS sector.
The string field $\Psi$ is defined on a subspace $\mathcal{H}_{c}$ of the worldsheet CFT Hilbert space annihilated by $L_{0}^{-} \equiv L_{0} - \bar{L}_{0}$ and $b_{0}^{-} \equiv b_{0} - \bar{b}_{0}$.
Moreover, the vectors in $\mathcal{H}_{c}$ are taken to be in the $(-1, -1)$ picture, which is canonical for the NS-NS vertex operators.
\subsection{SFT action}\label{sec:sftaction}
Suppose we have a basis $\{|\phi_{r}\rangle\}$ for $\mathcal{H}_{c}$, where $r$ denotes both discrete and continuous indices.
The string field can then be expanded as
\begin{equation}
\Psi = \sum_{r}\psi_{r}|\phi_{r}\rangle,
\end{equation}
where $\psi_{r}$ are spacetime fields including the massless graviton, Kalb-Ramond and dilaton field which will be of main interest in this paper.
For closed string theory, we take $\psi_{r}$ to have the same Grassmann parity as $|\phi_{r}\rangle$.
This is consistent with including Ramond sector fields as well.

The action $S[\Psi]$ takes the form of a cyclic $L_{\infty}$ action \cite{Zwiebach:1992ie}
\begin{equation}
  S[\Psi] = \frac{1}{2}\omega(\Psi, Q\Psi) + \sum_{n=2}^{\infty}\frac{1}{(n+1)!}\omega(\Psi, L_{n}(\Psi^{n})),
\end{equation}
where $\omega$ is the graded symmetric version of the BPZ inner product $\langle\cdot, \cdot\rangle$
\begin{equation}
\omega(\Psi_{1}, \Psi_{2}) \equiv {(-1)}^{\Psi_{1}} \langle \Psi_{1}, \Psi_{2} \rangle,
\end{equation}
and $L_{n}:\, \mathcal{H}_{c}^{n} \to \mathcal{H}_{c} \,(n \geq 2)$ are graded-antisymmetric multilinear maps defined via a rather complicated recursive procedure \cite{Erler2014}.
We also follow the convention that $L_{1}\equiv Q,$ the BRST-charge of the worldsheet theory.
For the classical Batalin-Vilkovisky action, only fields with zero spacetime ghost number are considered, which corresponds to worldsheet ghost number two.

The explicit construction of the products $L_{n}$ is given by a recursive procedure which determines the superstring products using the bosonic string products $l_{n}$.
We denote the picture number $(p, q)$ carried by the product as $L_{n}^{p, q}$.
In this notation, the product $L_{n}^{(n-1, n-1)} \equiv L_{n}$ is the product with desired picture number, since the tree level superstring amplitudes are non-vanishing only if the total picture number is $(-2, -2)$.
The products $L_{n}^{(0, 0)} = l_{n}$ with no extra picture number are identified with the bosonic string products which are known (see, e.g.~\cite{Zwiebach1991,Costello2022}).
The detailed construction is discussed in appendix~\ref{app:sft}.
For our purpose, only the first few string products are needed in detail.
However, for the sake of notational efficiency we briefly review the coalgebra notation, which conveniently expresses the $L_{\infty}$ relations and the recursive procedure.
Consider the graded symmetrized tensor algebra of the space of string fields $\mathcal{H}$
\begin{equation}
  S\mathcal{H} = \oplus_{n=0}^{\infty}\mathcal{H}^{\wedge n}.
\end{equation}
The wedge product is related to the tensor product through the formula
\begin{equation}
\Phi_{1}\wedge \Phi_{2} \wedge \dots \wedge \Phi_{n} = \sum_{\sigma} (-1)^{\epsilon(\sigma)} \Psi_{\sigma(1)} \otimes \Psi_{\sigma(2)} \otimes \dots \otimes \Psi_{\sigma(n)}, \quad \Phi_{i} \in \mathcal{H}.
\end{equation}
The sum is over all distinct permutations $\sigma$ of $1, \dots, n$, and the sign $(-1)^{\epsilon(\sigma)}$ is the obvious sign obtained by moving $\Phi_{(1)}, \Psi_{(2)}, \dots, \Phi_{n}$ past each other into the order prescribed by $\sigma$.
Any $n$-product $b_{n}: \mathcal{H}^{\wedge n} \to \mathcal{H}$ can be extended to a coderivation on the tensor coalgebra $\mathbf{b}_{n}: S\mathcal{H} \to S\mathcal{H} $, denoted by bold font letters, that is
\begin{equation}
  \mathbf{b}_{n} = \sum_{N=n}^{\infty} \left( b_{n} \wedge \mathbb{I}^{\otimes N-n} \right) P_{N},
\end{equation}
where $P_{N}$ is the projector to the $N$-fold symmetrized tensor space.
Given two coderivations $\mathbf{b}_{n}$ and $\mathbf{c}_{m}$, their commutator is another coderivation derived from a $(n+m-1)$-product
\begin{equation}
[ b_{m}, c_{n} ] \equiv b_{m} \left( c_{n} \wedge \mathbb{I}^{\otimes m - 1} \right) -(-1 )^{\mathrm{deg} ( b_{m} ) \mathrm{deg} ( c_{n} )} c_{n} \left( b_{m} \wedge \mathbb{I}^{\otimes n - 1} \right).
\end{equation}
With the coalgebra notation, the $L_{\infty}$ relations of the closed string products can be compactly written as
\begin{equation}
  [ \mathbf{L}_{1}, \mathbf{L}_{n} ]+[ \mathbf{L}_{2}, \mathbf{L}_{n-1} ]+\cdots+[ \mathbf{L}_{n-1}, \mathbf{L}_{2} ]+[ \mathbf{L}_{n}, \mathbf{L}_{1} ]=0,
\end{equation}
from which we can read off the relevant ones in the following:
\begin{equation}\label{eqn:l2}
[\mathbf{Q}, \mathbf{L}_{2}] = 0,
\end{equation}
and
\begin{equation}\label{eqn:l3}
[\mathbf{Q}, \mathbf{L}_{3}] + \frac{1}{2} [\mathbf{L}_{2}, \mathbf{L}_{2}] = 0.
\end{equation}

\subsection{Perturbative approach to exact marginality}\label{ssec:pa}

Conformal field theories are not isolated.
In a given CFT, operators are classified as relevant, irrelevant and marginal according to their conformal dimensions.
Among those the marginal operators of conformal dimension one are related to the possible deformations of a conformal field theory.
Such a deformation generated by $V_{i}(z)$ is of the form
\begin{equation}
\delta S \propto \int \mathrm{d}^{2}z \sum_{i} g_{i} V_{i}(z),
\end{equation}
where $g_{i}$'s are constants corresponding to coordinates tangent to the moduli space of CFTs.
The operator $V_{i}$ should be of conformal dimension $(1,1)$ and these operators are called the marginal operators.
However, the marginality of $V(z)$ may only preserve the classical conformal symmetry of the action.
Checking exact marginality usually breaks into an order-by-order procedure to see if the deformation preserves the conformal dimension of itself.
One should compute the exact beta-function to see if it vanishes.
Intuitively, we need to require the vanishing of integrals of the three-point functions
\begin{equation}
\left\langle V_{i}(z) V_{i}(w) \int \mathrm{d}^{2} z' V_{i}(z')  \right\rangle = 0
\end{equation}
to guarantee the two-point function remains unperturbed.
If all those perturbations, including higher order terms, vanish, then the perturbation is exactly marginal.
In general, it is difficult to verify by examination of $(n + 2)$-point functions that an operator remains marginal to all orders.

Studying the conformal manifold of two-dimensional conformal field theory is of particular interest as it describes the landscape of string theory vacua.
In such cases, this conformal manifold is endowed with a spacetime interpretation of the conformal field theory when we view it as a classical solution of string theory~\cite{Seiberg1988}.
Various works have been done in this direction.
However, it is difficult in general without an extended symmetry. For example, the $\mathcal{N} = 2$ superconformal field theories are fairly well understood.
One can identify all the exactly marginal fields, and, even better, find out that the conformal manifold has locally the structure of the product of the complexified Kähler moduli and the complex moduli.

The case with $\mathcal{N} = (1, 1)$ supersymmetry is, however, elusive.
It is known that string field theory provides a unified framework to treat different backgrounds as the classical solution to the string field equation of motion~\cite{Sen1994,Sen2018} and thus can be used to study our orbifold setup by finding higher order corrections to deformation of the twisted moduli.

We consider a perturbative solution
\begin{equation}\label{eqn:ps}
\Psi = \sum_{n=1}^{\infty} \lambda^{n} \Psi^{(n)}
\end{equation}
to the SFT equation of motion
\begin{equation}
Q\Psi + \sum_{n=2}^{\infty} \frac{1}{n!} L_{n}(\Psi^{n}) = 0,
\end{equation}
where $\lambda$ is the dimensionless perturbing parameter.
At each order of $\lambda$ we need to solve the linear equations
\begin{align}
Q\Psi^{(1)} &= 0,\\
Q\Psi^{(2)} &= -\frac{1}{2}L_{2}(\Psi^{(1)}, \Psi^{(1)}),\\
  Q\Psi^{(3)} &= -\frac{1}{2}L_{2}(\Psi^{(1)}, \Psi^{(2)}) - \frac{1}{2}L_{2}(\Psi^{(2)}, \Psi^{(1)}) - \frac{1}{6}L_{3}(\Psi^{(1)}, \Psi^{(1)}, \Psi^{(1)}),\\
  \nonumber &\cdots,
\end{align}
which require the RHS terms to be $Q$-exact.
It is easy to verify that the RHS terms are all $Q$-closed, thanks to the $L_{\infty}$ relations.
For example, using the first two relations (\ref{eqn:l2}) and (\ref{eqn:l3}) we can compute
\begin{align}
  \begin{split}
    &QL_{2}(\Psi^{(1)}, \Psi^{(1)}) \\
    &= -\mathbf{L}_{2}\mathbf{Q}(\Psi^{(1)}\wedge \Psi^{(1)})\\
    &= 0,
  \end{split}\\
  \begin{split}
    &Q\big( L_{2}(\Psi^{(1)}, \Psi^{(2)}) + L_{2}(\Psi^{(2)}, \Psi^{(1)}) + \frac{1}{3}L_{3}(\Psi^{(1)}, \Psi^{(1)}, \Psi^{(1)})\big) \\
    &= -\mathbf{L}_{2}\mathbf{Q}(\Psi^{(1)}\wedge \Psi^{(2)}) - \mathbf{L}_{2}\mathbf{Q}(\Psi^{(2)}\wedge \Psi^{(1)}) -\frac{1}{3}\mathbf{L}_{3}\mathbf{Q}(\Psi^{(1)}\wedge \Psi^{(1)}\wedge \Psi^{(1)}) - \frac{1}{3}\mathbf{L}_{2}\mathbf{L}_{2}(\Psi^{(1)}\wedge \Psi^{(1)}\wedge \Psi^{(1)}) \\
    &= L_{2}(\Psi^{(1)}, L_{2}(\Psi^{(1)}, \Psi^{(1)})) + L_{2}(L_{2}(\Psi^{(1)}, \Psi^{(1)}), \Psi^{(1)}) - \frac{1}{3}\mathbf{L}_{2}\mathbf{L}_{2}(\Psi^{(1)}\wedge \Psi^{(1)}\wedge \Psi^{(1)})\\
    &= 0,
  \end{split}
\end{align}
where $\Psi^{(1)}$ and $\Psi^{(2)}$ are assumed to satisfy the perturbative equation of motion.
Note that acting $\mathbf{L}_{2}$ on $\Psi^{(1)} \wedge \Psi^{(1)} \wedge \Psi^{(1)}$ gives a permutation factor $3!$.
Therefore, the $Q$-exactness requirement is met if the RHS terms are mapped into the trivial class by projecting to the cohomology $H(Q)$.
Furthermore, at the cohomological level, it is possible to restrict all states to the kernel of $L_{0}^{+} \equiv L_{0} + \bar{L}_{0}$, since we can construct a contracting homotopy $\Delta = b^{+}_{0}/L_{0}^{+}$, where $b_{0}^{+}\equiv b_{0} + \bar{b}_{0}$. 
On the complement of $\mathrm{ker}(L_{0}^{+})$, $\Delta$ is well-defined and thus satisfies $\left\{Q, \Delta\right\} = 1$. 
Therefore, the contracting homotopy $\Delta$ trivializes the cohomology except for those states in $\mathrm{ker}(L_{0}^{+})$.
We denote the compositional projector $P_{0}$ and hence we obtain the cohomological obstructions up to the third order,
\begin{align}\label{eq:O23}
\mathcal{O}^{(2)} &= P_{0}L_{2}(\Psi^{(1)}, \Psi^{(1)}), \\
\mathcal{O}^{(3)} &= P_{0}\bigg( L_{2}\left(\Psi^{(1)}, \Psi^{(2)}\right) + L_{2}\left(\Psi^{(2)}, \Psi^{(1)}\right) + \frac{1}{3} L_{3}\left(\Psi^{(1)}, \Psi^{(1)}, \Psi^{(1)}\right) \bigg),
\end{align}
and the perturbative solution
\begin{equation}
\Psi^{(2)} = -\frac{1}{2}Q^{-1}L_{2}\left(\Psi^{(1)}, \Psi^{(1)}\right),
\end{equation}
with the explicit expression for $L_2$ given in eqn. (5.4) of~\cite{Erler2014}.

\section{Resolving the orbifold singularity}\label{sec:res}

A marginal deformation in the worldsheet CFT is exactly marginal if the corresponding solution of the linearized equation of motion of SFT can be integrated to a solution of the nonlinear equation of motion.
We begin by expanding the field in a perturbation series, as in (\ref{eqn:ps}), with dimensionless parameter $\lambda$ and
\begin{equation}
\lambda \Psi^{(1)} = \mathcal{V}
\end{equation}
which, for $\mathcal{V}$ as in (\ref{eqn:md}), is a solution of the linearized equation of motion $Q\mathcal{V} = 0$ that describes an infinitesimal blow-up of the orbifold singularity.
We then, as analyzed in subsection~\ref{ssec:pa}, have an infinite series of string field corrections $\Psi^{(2)}, \dots$ and cohomological obstructions $\mathcal{O}^{(2)}, \mathcal{O}^{(3)}$ to solving the equations.

\subsection{Second order obstruction}\label{subsec:secob}

Starting from~\eqref{eq:O23} and using the explicit expression for $L_2$ in~\cite{Erler2014} one can show that the obstruction $\mathcal{O}^{(2)}$ factorizes as\footnote{As in~\cite{Erler2014}, we drop a $b_0^-$ in the definition of the string products since we focus on the equations of motion.} 
 \begin{equation}\label{eq:pl2}
 \begin{split}
 \mathcal{O}^{(2)} =& P_{0}\left[ \ell_{2}\bigg( \mathcal{X}\left( cVe^{-\phi} \right), cVe^{-\phi} \bigg) + \ell_{2} \bigg( cVe^{-\phi}, \mathcal{X} \left( cVe^{-\phi} \right) \bigg) +\mathcal{X} \ell_{2}\left( cVe^{-\phi}, cVe^{-\phi} \right) \right] \otimes\;\text{anti-hol. part}\\
 \equiv& P_{0} \left[(X\circ \ell_2 )\bigg(  cVe^{-\phi} , cVe^{-\phi} \bigg)\right]\otimes\;\text{anti-hol. part}
,
\end{split}
\end{equation}
where $\ell_2$ (and $\bar\ell_2$) are the holomorphic and anti-holomorphic parts of the bosonic string products introduced in section~\ref{sec:sftaction}.
The following calculation follows closely that of~\cite{Mattiello2019} for the open string.
Consider first the term in~\eqref{eq:pl2},
\begin{equation}
  \begin{split}
    &P_{0}\left[ \ell_{2}\bigg( \mathcal{X}\left( cVe^{-\phi} \right), cVe^{-\phi} \bigg) + \ell_{2} \bigg( cVe^{-\phi}, \mathcal{X} \left( cVe^{-\phi} \right) \bigg) \right] \\
    &= \lim_{z \to 0} P_{0} \left[ \mathcal{X}\left( cVe^{-\phi} \right)(z) cVe^{-\phi}(0) + cVe^{-\phi}(z) \mathcal{X}\left( cVe^{-\phi} \right) (0) \right],
  \end{split}
\end{equation}
where the action of the picture-changing operator $\mathcal{X}$ on $cVe^{-\phi}$ can be computed from the OPE, yielding
\begin{equation}
  \mathcal{X}\left( cVe^{-\phi} \right) = -cV_{1} + \frac{1}{4}\gamma V,
\end{equation}
where $V_{1}$ denotes a matter primary field of conformal dimension one.
Since the OPE $V(z)V(0)$ contains a single pole and the OPE $V(z)V_{1}(0)$ does not, we conclude that
\begin{equation}
  \begin{split}
    &P_{0}\left[ \ell_{2}\bigg( \mathcal{X}\left( cVe^{-\phi} \right), cVe^{-\phi} \bigg) + \ell_{2} \bigg( cVe^{-\phi}, \mathcal{X} \left( cVe^{-\phi} \right) \bigg) \right] \\
    &= \lim_{z \to 0} P_{0} \left[ \left( -cV_{1} + \frac{1}{4}\gamma V \right)(z) cVe^{-\phi}(0) + cVe^{-\phi}(z) \left( -cV_{1} + \frac{1}{4}\gamma V \right) (0) \right] \\
    &= \lim_{z \to 0} P_{0} \left[ \frac{1}{4}zc\eta(0) \bigg( V(z)V(0) - V(z)V(0) \bigg) \right],
  \end{split}
\end{equation}
where commuting $c$ with $V$ picks up a minus sign.
We then consider the remaining term
\begin{equation}
P_{0}\left[ \mathcal{X} \ell_{2}\left( cVe^{-\phi}, cVe^{-\phi} \right) \right] = \mathcal{X} P_{0}\left[ \ell_{2}\left( cVe^{-\phi}, cVe^{-\phi} \right) \right].
\end{equation}
The projector part can be computed from OPE
\begin{equation}\label{eqn:mvv}
P_{0}\left[ \ell_{2}\left( cVe^{-\phi}, cVe^{-\phi} \right) \right] = \lim_{z \to 0} P_{0} \left[ cVe^{-\phi}(z) cVe^{-\phi}(0) \right] = \partial \left(c\partial{c}e^{-2\phi}V'_{0}\right) + c\partial{c} V'_{1}e^{-2\phi},
\end{equation}
where $V'_{0}$ and $V'_{1}$ are matter vertex operators of conformal dimension zero and one, respectively.
The first term in (\ref{eqn:mvv}) is BRST-exact, i.e.
\begin{equation}
\partial \left(c\partial{c}e^{-2\phi}V'_{0}\right) = Q \left(\partial{c}e^{-2\phi}V'_{0}\right)
\end{equation}
and projects to zero cohomology by $P_{0}$, and thus
\begin{equation}
  \mathcal{X}P_{0}\left[ \ell_{2}\left( cVe^{-\phi}, cVe^{-\phi} \right) \right] = P_{0}\left[ \mathcal{X} \left( c\partial{c} V'_{1}e^{-2\phi} \right) \right].
\end{equation}
Note that in reaching this conclusion we did not need to resort to the large Hilbert space.
For the remaining term, we note that, going into the large Hilbert space we have
\begin{equation}
  \mathcal{X} \ell_{2}\left( cVe^{-\phi}, cVe^{-\phi} \right) = \left\{ Q, \xi \right\} \left( c\partial{c} V'_{1}e^{-2\phi} \right) = Q \xi\left( c\partial{c} V'_{1}e^{-2\phi} \right),
\end{equation}
where the term $\xi(\cdots)$ will have a vanishing inner product with states in the large Hilbert space due to the presence of an extra $\xi$-zero mode there.
The BRST charge extracts the double pole of the OPE $T_{F}(z)V'_{1}(0)$, while the OPE\footnote{See eqn (7.13) in~\cite{Mattiello2019}.}
\begin{equation}
T_{F}(z)\psi^{ij}(0) \sim \frac{1}{z}\left( \partial{X}^i\psi^j - \partial{X}^i\psi^{i} \right)
\end{equation}
has only a single pole.
Therefore we conclude that the second order obstruction vanishes, and the perturbative solution
\begin{equation} \label{eq:qp2}
\Psi^{(2)} = -\frac{1}{2}Q^{-1} L_{2}^{(1, 1)} (\Psi^{(1)}, \Psi^{(1)})
\end{equation}
is well-defined.

\subsection{Gravitational instanton from the string field}\label{subsec:gra}
We can extract the spacetime field contents from the string field theory solution.
Consider the vertex operator in the $(-1,-1)$ picture corresponding to the spacetime 2-tensor $W_{ij}$ 
\begin{equation}\label{eq:V_W}
  V_{W} = \int \frac{\mathrm{d}^{4} k}{{(2\pi)}^{2}} W_{ij}(k)c \psi^{i}  e^{-\phi} \bar{c} \bar{\psi}^{j}  e^{-\bar{\phi}}(\bar{z})e^{-i k {X}}(z,\bar z),
\end{equation}
where $k$ is the four-dimensional Euclidean momentum.
It is straightforward to compute its BRST variation 
\begin{equation}\label{eqn:qvw}
  \begin{split}
    QV_{W} = &\int \frac{\mathrm{d}^{4} k}{{(2\pi)}^{2}} \bigg( -\frac{k^{2}}{2} W_{ij}(k)c\partial{c} \psi^{i} e^{-i k X}(z) e^{-\phi} \bar{c}\bar{\partial}\bar{c} \bar{\psi}^{j} e^{-i k \bar{X}} e^{-\bar{\phi}}(\bar{z}) \\
    &+ \frac{ik^{i}}{4} W_{ij}c\eta e^{-i k X}(z)\bar{c} \bar{\psi}^{j} e^{-i k \bar{X}} e^{-\bar{\phi}}(\bar{z}) + \frac{ik^{j}}{4} W_{ij}c\psi^{i} e^{-i k X}(z)\bar{c} \bar{\eta} e^{-i k \bar{X}}(\bar{z}) \bigg).
  \end{split}
\end{equation}
Then, we see from \eqref{eq:qp2} that we can extract $W_{ij}$ by taking the inner product with $-\frac{1}{2}L_{2}^{(1, 1)} (\Psi^{(1)}, \Psi^{(1)})$, i.e.
\begin{equation}
-\frac{k^{2}}{2} W_{ij}(k) = \left\langle c \psi_{i} e^{-i k X} e^{-\phi} \bar{c} \bar{\psi}_{j} e^{-i k \bar{X}} e^{-\bar{\phi}}, -\frac{1}{2} L_{2}^{(1, 1)} (\Psi^{(1)}, \Psi^{(1)}) \right\rangle.
\end{equation}
Moreover, the effect of assigned picture-changing operator in $L_{2}^{(1, 1)}$ can be moved around, since $\mathcal{X} = \left\{ Q, \xi \right\}, \bar{\mathcal{X}} = \left\{ Q, \bar{\xi} \right\}$ and none of the vertex operators here involve the $\eta$ ghost, we can move $\mathcal{X}\bar{\mathcal{X}}$ from either input to the first operator in spite of it being off-shell, resulting in a $(0, 0)$ picture number (see, e.g. section 6 of  \cite{Erler:2013xta} for a detailed discussion),
\begin{equation}\label{eq:fvp}
k^{2} W^{(2)}_{ij}(k) = \left\langle \left( f_{\infty} \circ \Psi^{(1)} \right) (\infty) \left( f_{1} \circ V_{ij}^{(0, 0)} \right) (0) \left( f_{0} \circ \Psi^{(1)} \right) (0) \right\rangle,
\end{equation}
where the operator in natural picture number
\begin{equation} 
V_{ij}^{(-1, -1)} = c \psi_{i} e^{-i k X} e^{-\phi} \bar{c} \bar{\psi}_{j} e^{-i k \bar{X}} e^{-\bar{\phi}}
\end{equation}
needs to be assigned the picture-changing operator $\mathcal{X} \bar{\mathcal{X}}$ and turns into
\begin{equation}\label{eqn:metop}
V_{ij}^{(0, 0)} = c \left( i \partial X_{i} - \frac{\alpha'}{2} (k \psi) \psi_{i} \right) e^{-i k X} \bar{c} \left( i \bar{\partial} \bar{X}_{j} - \frac{\alpha'}{2} (k \bar{\psi}) \bar{\psi}_{j} \right) e^{-i k \bar{X}},
\end{equation}
and $f_{0}$, $f_{1}$, and $f_{\infty}$ are the three conformal maps that are needed to construct the bosonic string product $l_{2}$, see appendix~\ref{app:sft} for details. If one were to assume that all insertions are on-shell, eqn. \eqref{eq:fvp} should be familiar from perturbative string theory.  
Note that in the following calculation we may drop the terms containing the boson current since they give contributions proportional to $k_{i}k_{j}$ and are thus infinitesimally diffeomorphic to zero.
So there are only the contributions due to the non-vanishing momentum $k^{i}$.
We note here that the action of a conformal map $f_{z_{0}}$ on a primary operator $O(z)$ with conformal dimension $h$ is
\begin{equation} 
f_{z_{0}} \circ O(z) = f'_{z_{0}} (z)^{h} O(f_{z_{0}} (z)),
\end{equation}
where $z_{0}$ is the image of the origin.
For the two-product we use three maps $f_{0}, f_{1}$ and $f_{\infty}$, however the only non-trivial action is $f_{1}$ on the operator.
Therefore the correlation function reduces to
\begin{equation} 
W^{(2)}_{ij}(k) = \frac{1}{k^{2}}\left| f_{1}'(0) \right|^{\alpha' k^{2}} \left< V^{(-1)} \bar{V}^{(-1)} (\infty) V_{ij}^{(0, 0)}(1) V^{(-1)} \bar{V}^{(-1)} (0) \right>.
\end{equation}
In the holomorphic sector, it can be split into four sub-amplitudes with generic vertex operator insertion, which belong to different CFTs and are independent from each other:
\begin{equation}
    \begin{split}
    w_{\alpha} w_{\beta} k^{\mu} & \left\langle c(z_{1}) c(z_{2}) c(z_{3}) \right\rangle \left\langle e^{-\phi(z_{1})} e^{-\phi(z_{3})} \right\rangle \\
    & \left\langle \Delta(z_{1}) e^{-i k \cdot X}(z_{2}) \Delta(z_{3}) \right\rangle \left\langle S^{\dot{\alpha}} \psi_{i} \psi_{j}(z_{2}) S^{\dot{\beta}}(z_{3}) \right\rangle.
    \end{split}
\end{equation}
All the correlation functions are known and thus we can calculate
\begin{equation} 
W^{(2)}_{ij}(k) = \frac{\left| f_{1}'(0) \right|^{\alpha' k^{2}}}{16k^{2}} \left( \frac{z_{13}}{z_{12} z_{23}} \cdot \frac{\bar{z}_{13}}{\bar{z}_{12} \bar{z}_{23}} \right)^{\alpha' k^{2} / 2} k^{k}k^{l} w_{\alpha} (\sigma_{ki})^{\alpha \beta} w_{\beta} \bar{w}_{\gamma} (\sigma_{lj})^{\gamma \delta} \bar{w}_{\delta} e^{-i k \cdot x_{0} - i k \cdot \bar{x}_{0}},
\end{equation}
where $\sigma_{ij}$ (not to be confused with the twist field $\sigma^{n}$) are the $SO(4)$ generators that are defined in terms of Pauli matrices $(\tau^{1}, \tau^{2}, \tau^{3})$ in the following way:
\begin{equation}
\sigma_{ij} = \frac{1}{2} \left( \sigma_{i}\bar{\sigma}_{j} - \sigma_{j}\bar{\sigma}_{j} \right),
\end{equation}
where in the Euclidean space
\begin{equation}
\sigma_{i} = (\mathbb{I}, -i\tau^{1}, -i\tau^{2}, -i\tau^{3}) \quad \mathrm{and} \quad \bar{\sigma}_{i} = (\mathbb{I}, i\tau^{1}, i\tau^{2}, i\tau^{3}).
\end{equation}
Here $x_{0}$ and $\bar{x}_{0}$ denote the zero-modes of $X(z)$ and $\bar{X}(\bar{z})$ respectively, which correspond to the position of the orbifold singular point that is chosen to be zero in the following.
After setting $z_{1} \to \infty, z_{2} = 1$ and $z_{3} = 0$ we end up with
\begin{equation}\label{eqn:metric}
  W_{ij}^{(2)} = \frac{\left| f_{1}'(0) \right|^{\alpha' k^{2}}}{16k^{2}} k^{k} k^{l} w_{\alpha} (\sigma_{ki})^{\alpha\beta} w_{\beta} \bar{w}_{\gamma} (\sigma_{lj})^{\gamma\delta} \bar{w}_{\delta}.
\end{equation}
The map $f_{1}$ is explicitly given by the composition $f_{1}(z) = G(g_{1}(z))$ with
\begin{equation} 
g_{1}(z) = \left( \frac{i - z}{i + z} \right)^{2/3}
\end{equation}
and
\begin{equation} 
G(z) = \frac{(1 - e^{2\pi i / 3}) (z - e^{-2 \pi i / 3})}{(1 - e^{- 2 \pi i / 3}) (z - e^{2 \pi i / 3})},
\end{equation}
so its derivative at the origin is
\begin{equation} 
f'_{1}(0) = \frac{4}{3 \sqrt{3}},
\end{equation}
which is suppressed by the power $\alpha' k^{2}$ to $1$ once we take the field theory limit $\alpha' k^{2} \ll 1$.
The spacetime field $W_{ij}$ is decomposed as\footnote{More generally,  the massless part of the string field has the form 
$V_W=c\tilde c\left(W_{ij}(x)\,\psi^i\tilde\psi^j+\phi(x)\,\gamma\tilde\beta+\tilde\phi(x)\,\tilde\gamma\beta+\cdots\right)$, where the dilaton is identified with a certain linear combination of $W^i_i$, $\phi$ and $\tilde\phi$. Our vertex operator \eqref{eq:V_W} then corresponds to the gauge where $\tilde\phi=\phi=0$. See \cite{Bonezzi:2020jjq} for the details.} 
\begin{equation}
W_{ij} = G_{ij} + B_{ij} + \Phi \delta_{ij},
\end{equation}
where $G$, $B$, and $\Phi$ are the traceless symmetric, anti-symmetric, and trace parts, respectively.
We now examine the traceless symmetric part of (\ref{eqn:metric}) that represents a perturbation to the metric.
The Fourier transform gives
\begin{equation}\label{eqn:ft}
  \int \frac{\mathrm{d}^{4}k}{(2\pi)^{2}} \frac{k_{i}k_{j}}{k^{2}} e^{ikx} = -\partial_{i}\partial_{j}\int \frac{\mathrm{d}^{4}k}{(2\pi)^{2}} \frac{1}{k^{2}} e^{ikx} = -\partial_{i}\partial_{j}(\frac{1}{x^{2}})  = -2(\frac{\delta_{ij}}{x^{4}} - \frac{4x_{i}x_{j}}{x^{6}}).
\end{equation}
We would like to comment that even though we start with the flat metric on $\mathbb{C}^{2} / \mathbb{Z}_{2}$, which is Kähler, the full metric solution to the blowing-up equation is not necessarily so since we are considering marginal deformation of $\mathcal{N} = (1,1)$ worldsheet theory.
To begin with, we take $w_{\alpha} = \bar{w}_{\alpha} = (\frac{1}{\sqrt{2}}, \frac{1}{\sqrt{2}})$; we then have

\begin{equation} 
G_{ij}^{(2)} \sim
\begin{pmatrix}
k^{2} - 4k_{2}^{2} & 4k_{1}k_{2} & -4k_{2}k_{4} & 4k_{2}k_{3}\\
4k_{1}k_{2} & k^{2} - 4k_{1}^2 & 4k_{1}k_{4} & -4k_{1}k_{3}\\
-4k_{2}k_{4} & 4k_{1}k_{4} & k^{2} - 4k_{4}^{2} & 4k_{3}k_{4}\\
4k_{2}k_{3} & -4k_{1}k_{3} & 4k_{3}k_{4} & k^{2} - 4k_{3}^{2}
\end{pmatrix}/k^{2},
\end{equation}
and the dilaton
\begin{equation}
\Phi^{(2)} \sim 1,
\end{equation}
which is localized at the origin in the coordinate space.
Consequently, in the region $x\neq 0$, the dilaton remains constant, ensuring that the metric satisfies the Ricci-flat condition $R_{ij} = 0$ at this order, consistent with the Eguchi-Hanson vacuum solution.
We note that the Kalb-Ramond field $B_{ij}$ vanishes,
\begin{equation}
B_{ij}^{(2)} \sim 0
\end{equation}
with this choice of moduli.
Going back to the coordinate space using (\ref{eqn:ft}) we have
\begin{equation}\label{eq:gbc}
G_{ij}^{(2)}(x) \sim
\begin{pmatrix}
-x^{2} + 4x_{2}^{2} & -4x_{1}x_{2} & 4x_{2}x_{4} & -4x_{2}x_{3}\\
-4x_{1}x_{2} & -x^{2} + 4x_{1}^2 & -4x_{1}x_{4} & 4x_{1}x_{3}\\
4x_{2}x_{4} & -4x_{1}x_{4} & -x^{2} + 4x_{4}^{2} & -4x_{3}x_{4}\\
-4x_{2}x_{3} & 4x_{1}x_{3} & -4x_{3}x_{4} & -x^{2} + 4x_{3}^{2}
\end{pmatrix}/x^{6}.
\end{equation}
An infinitesimal diffeomorphism generated by the vector field $(x_{1},x_{2},x_{3},x_{4})/x^{2}$ brings the metric to
\begin{equation}
G_{ij}^{(2)}(x) \sim
\begin{pmatrix}
x_{1}^{2} + x_{2}^2 - x_{3}^{2} - x_{4}^{2} & 0 & 2(x_{1}x_{3}+x_{2}x_{4}) & 2(x_{1}x_{4}-x_{2}x_{3})\\
0 & x_{1}^{2} + x_{2}^2 - x_{3}^{2} - x_{4}^{2} & -2(x_{1}x_{4}-x_{2}x_{3}) & 2(x_{1}x_{3}+x_{2}x_{4})\\
2(x_{1}x_{3}+x_{2}x_{4}) & -2(x_{1}x_{4}-x_{2}x_{3}) & -x_{1}^{2} - x_{2}^2 + x_{3}^{2} + x_{4}^{2} & 0\\
2(x_{1}x_{4}-x_{2}x_{3}) & 2(x_{1}x_{3}+x_{2}x_{4}) & 0 & -x_{1}^{2} - x_{2}^2 + x_{3}^{2} + x_{4}^{2}
\end{pmatrix}/x^{6}.
\end{equation}
Finally we consider $(x_{1},x_{2},x_{3},x_{4})\mapsto(-x_{4},-x_{3},x_{2},x_{1})$ to rearrange the coordinates and we obtain the metric
\begin{equation}
G_{ij}^{(2)}(x) \sim
\begin{pmatrix}
-x_{1}^{2} - x_{2}^2 + x_{3}^{2} + x_{4}^{2} & 0 & -2(x_{1}x_{3}+x_{2}x_{4}) & -2(x_{1}x_{4}-x_{2}x_{3})\\
0 & -x_{1}^{2} - x_{2}^2 + x_{3}^{2} + x_{4}^{2} & 2(x_{1}x_{4}-x_{2}x_{3}) & -2(x_{1}x_{3}+x_{2}x_{4})\\
-2(x_{1}x_{3}+x_{2}x_{4}) & 2(x_{1}x_{4}-x_{2}x_{3}) & x_{1}^{2} + x_{2}^2 - x_{3}^{2} - x_{4}^{2} & 0\\
-2(x_{1}x_{4}-x_{2}x_{3}) & -2(x_{1}x_{3}+x_{2}x_{4}) & 0 & x_{1}^{2} + x_{2}^2 - x_{3}^{2} - x_{4}^{2}
\end{pmatrix}/x^{6},
\end{equation}
which is exactly the real form of the Eguchi-Hanson metric
\begin{equation}\label{eq:SEH}
  H_{I\bar{J}} = {(1+\frac{\rho^{4}}{x^{4}})}^{1/2}
  \begin{pmatrix} 1-\frac{\rho^{4} \bar{z}_{1} z_{1}}{x^2(\rho^4+x^4)} & -\frac{\rho^4 \bar{z}_{1}z_{2}}{x^2(\rho^4+x^4)} \\
    -\frac{\rho^4 \bar{z}_{2} z_{1}}{x^2(\rho^4+x^4)} & 1 -\frac{\rho^4 \bar{z}_{2} z_{2}}{x^2(\rho^4+x^4)} \end{pmatrix}
  = \delta_{I\bar{J}} +
  \begin{pmatrix}
    \frac{1}{2x^{4}}-\frac{\bar{z}_{1} z_{1}}{x^6} & -\frac{\bar{z}_{1} z_{2}}{x^6} \\
    -\frac{\bar{z}_{2}z_{1}}{x^6} & \frac{1}{2x^{4}} -\frac{\bar{z}_{2}z_{2}}{x^6}
  \end{pmatrix}\rho^4
 + O(\rho^{6}),
\end{equation}
where the complex structure is given by $z_{1} = x_{1} + ix_{2},\,z_{2} = x_{3} + ix_{4}$ and $\rho$ is the typical length scale of the instanton.  Since \eqref{eq:SEH} is hermitian in the new coordinates, we can read off the deformation of the complex structure from the corresponding real coordinate transformation below \eqref{eq:gbc}. At first order the Beltrami differential is then
\begin{align}
 \xi(\rho)=\frac{2\rho^2 z_i z_k}{|z|^4}   \partial_{ z_k}\mathrm{d}\bar z^i=-\rho^2 z_k\partial_{ z_k}\bar\partial(\frac{1}{|z|^2}).
\end{align}

It is worth noting that the Eguchi-Hanson metric is, by definition, hyperk\"ahler.
A general choice of moduli gives rise to a non-zero Kalb-Ramond field $B_{ij}$.
One would expect that such a metric profile is neither K\"ahler nor hyperk\"ahler.
Since the dilaton does not contribute to the Einstein equation away from the origin, and the Kalb-Ramond field only contributes a higher order term, we know that the metric satisfies the Ricci-flatness condition at this order.
The K\"ahlerness is, however, harder to show since it is not clear how to find a compatible complex structure deformed from $\mathbb{C}^{2}$.
However, in dimension four any Ricci-flat K\"ahler manifold must be hyperk\"ahler, which can be shown without referring to the complex structure.
Remarkably, we can show by computation that the Weyl tensor $C$ is self-dual (details are in appendix~\ref{app:weyl}).
\begin{equation}
C = C^{+},\quad C^{-} = 0,
\end{equation}
and we thus conclude that away from the origin the metric is K\"ahler to this order. 

 It is also possible to include a Kalb-Ramond field $B_{ij}$. For instance, for $w_{\alpha} =  (\frac{1}{\sqrt{2}}, \frac{1}{\sqrt{2}})(1+\frac{\epsilon}{2})$, $\bar w_\alpha= (\frac{1}{\sqrt{2}}, \frac{1}{\sqrt{2}})(1-\frac{\epsilon}{2})$, $\epsilon \ll 1$, we have the form \eqref{eqn:profile} with $u=\bar u=\frac{1}{2}+O(\epsilon^2)$ and $v=-\bar v= \epsilon+O(\epsilon^2)$,
 \begin{equation}\begin{split}
  B_{ij}^{(2)}&= -\epsilon\frac{16\mathcal{N}\left( \alpha' \right)}{x^{6}} \\
    &\times
\begin{pmatrix}0
 & -2{} (x_3x_2+x_4x_1)& -{} \left(x^2-2 x_{2}^2-2 x_{4}^2\right) & 2{} ( x_{2}x_{1}-x_{4}x_{3}) 
  \\
 2{} (x_3x_2+x_4x_1) &0  &-2{} \left(x_1 x_2-x_3x_4\right)  & {}(x^2-2x_1^2-2x_3^2)    \\
{} \left(x^2-2 x_{2}^2-2 x_{4}^2\right)  & 2{} ( x_{2}x_{1}-x_{4}x_{3})  & 0  & 2{}(x_2x_3+x_4x_1) 
  \\
 -2{} ( x_{2}x_{1}-x_{4}x_{3}) & -{}(x^2-2x_1^2-2x_3^2) &-2{}(x_2x_3+x_4x_1) &0  
\end{pmatrix}+O(\epsilon^2)
  \end{split},
\end{equation}
which turns out to be exact, 
\begin{align}
    B=\mathrm{d} A,\qquad A=-\epsilon\frac{ 8\mathcal{N}}{x^4}\begin{pmatrix}
     -x_3\\x_4\\x_1\\-x_2  
    \end{pmatrix}.
\end{align}
This then describes an asymptotic Eguchi-Hanson space with an exact Kalb-Ramond field.

\subsection{Third order obstruction}

The obstruction
\begin{equation}\label{eq:o3}
\begin{split}
  \mathcal{O}^{(3)} = \frac{1}{3}P_{0}&\bigg( 3L_{2}\left(\Psi^{(1)}, \Psi^{(2)}\right) + 3L_{2}\left(\Psi^{(2)}, \Psi^{(1)}\right) + L_{3}\left(\Psi^{(1)}, \Psi^{(1)}, \Psi^{(1)}\right) \bigg) \\
  = \frac{1}{3}P_{0}&\Bigg( -\frac{3}{2}L_{2}\bigg( \Psi^{(1)}, Q^{-1}L_{2}\left( \Psi^{(1)}, \Psi^{(1)} \right) \bigg) -\frac{3}{2}L_{2}\bigg( Q^{-1}L_{2}\left( \Psi^{(1)}, \Psi^{(1)} \right), \Psi^{(1)}\bigg) \\
 &+ L_{3}\left(\Psi^{(1)}, \Psi^{(1)}, \Psi^{(1)}\right) \Bigg) \\
  = \frac{1}{3}P_{0}&\Bigg( -3L_{2}\bigg( \Psi^{(1)}, Q^{-1}L_{2}\left( \Psi^{(1)}, \Psi^{(1)} \right) \bigg) + L_{3}\left(\Psi^{(1)}, \Psi^{(1)}, \Psi^{(1)}\right) \Bigg)
\end{split}
\end{equation}
is spanned on a basis $\left\{e_{i}\right\}$ of $H(Q)$ using the BPZ inner product
\begin{equation}
    \mathcal{O}^{(3)} = \sum_{i} e_{i}\left\langle e^{i}, \mathcal{O}^{(3)} \right\rangle = \frac{1}{3}\sum_{i} e_{i}\mathcal{O}^{(3)}_{i} ,
\end{equation}
where $\left\{e^{i}\right\}$ is the dual basis satisfying $\left\langle e^{i}, e_{j} \right\rangle = \delta^{i}_{j}$.

To show the vanishing of $\mathcal{O}^{(3)}$, we follow an approach similar to that of~\cite{Maccaferri:2019ogq}, whose essential logic is as follows: using $\bar{\mathcal{X}} = \{Q, \bar{\xi}\}$ and the fact that $\bar{\xi}$ zero mode must be saturated in the large Hilbert space, we can move the picture changing operator arbitrarily.
But whenever $Q$ hits a $Q^{-1}$, we get a $1 - P_{0}$ from the homotopy relation
\begin{equation}
QQ^{-1} + Q^{-1}Q = 1 - P_{0},
\end{equation}
representing a contact term and a boundary term in the moduli space, respectively.
The contact term cancels due to $L_{\infty}$ relations and we are left with propagator terms with PCOs moved, and boundary terms.
One further makes the observation that the vertices can be decomposed according to an additional $U(1)$ charge on the worldsheet,
\begin{equation}
\Psi = \Psi_{+} + \Psi_{-},
\end{equation}
and the total charge of the vertices has to be neutral to have a non-zero correlator.
Acting with the PCO on the vertex gives two terms: one with the same $U(1)$ charge but wrong ghost number, and one with neutral $U(1)$ charge and correct ghost number.
Therefore, the two picture-unchanged vertices must obtain contributions only from opposite $U(1)$ charge components.
Since we can choose the picture-changed vertices freely (up to boundary terms) in the propagator term, this puts inconsistent constraints on the $U(1)$ charge.
The propagator term then must vanish, and the obstruction is localized on boundary terms with $P_{0}$.
The story of Type II theory is different in that the presence of the holomorphic sector makes the $P_{0}$ terms vanish, too, as shown in subsection~\ref{subsec:secob}, and hence the obstruction vanishes.

We start with using the definition $\mathbf{L}_{2} = \left[ \mathbf{Q}, \boldsymbol{\lambda}_{2}^{(1,1)} \right]$.
When acting on two inputs, it gives $L_{2} = Q\lambda_{2}^{(1,1)} - \lambda_{2}^{(1,1)}\left( \mathbb{I} \otimes Q + Q \otimes \mathbb{I} \right)$ to get
\begin{equation}
\begin{split}
  2&L_{2}\bigg( \Psi^{(1)}, Q^{-1}L_{2}\left( \Psi^{(1)}, \Psi^{(1)} \right) \bigg) = -\lambda_{2}^{(1,1)}\bigg( Q\Psi^{(1)}, Q^{-1}L_{2}\left( \Psi^{(1)}, \Psi^{(1)} \right) \bigg) \\
   &- \lambda_{2}^{(1,1)}\bigg( \Psi^{(1)}, QQ^{-1}L_{2}\left( \Psi^{(1)}, \Psi^{(1)} \right) \bigg) + L_{2}\bigg( \Psi^{(1)}, Q^{-1}Q\lambda_{2}^{(1,1)}\left( \Psi^{(1)}, \Psi^{(1)} \right) \bigg) \\
   &- L_{2}\bigg( \Psi^{(1)}, Q^{-1}\lambda_{2}^{(1,1)}\left( Q\Psi^{(1)}, \Psi^{(1)} \right) \bigg) - L_{2}\bigg( \Psi^{(1)}, Q^{-1}\lambda_{2}^{(1,1)}\left( \Psi^{(1)}, Q\Psi^{(1)} \right) \bigg) \\
   &+ Q\lambda_{2}^{(1,1)}\bigg( \Psi^{(1)}, Q^{-1}L_{2}\left( \Psi^{(1)}, \Psi^{(1)} \right) \bigg),
\end{split}
\end{equation}
where the first, the fourth and the fifth terms vanish due to the on-shell condition and we will omit similar terms in the following.
The remaining terms are
\begin{equation}\label{eqn:obs2}
\begin{split}
  &-\lambda_{2}^{(1,1)}\bigg( \Psi^{(1)}, QQ^{-1}L_{2}\left( \Psi^{(1)}, \Psi^{(1)} \right) \bigg) + L_{2}\bigg( \Psi^{(1)}, Q^{-1}Q\lambda_{2}^{(1,1)}\left( \Psi^{(1)}, \Psi^{(1)} \right) \bigg) \\
  &+ Q\lambda_{2}^{(1,1)}\bigg( \Psi^{(1)}, Q^{-1}L_{2}\left( \Psi^{(1)}, \Psi^{(1)} \right) \bigg) \\
  &= -\lambda_{2}^{(1,1)}\bigg( \Psi^{(1)}, \left( 1 - P_{0} - Q^{-1}Q \right)L_{2}\left( \Psi^{(1)}, \Psi^{(1)} \right) \bigg) + L_{2}\bigg( \Psi^{(1)}, \left( 1 - P_{0} - QQ^{-1} \right)\lambda_{2}^{(1,1)}\left( \Psi^{(1)}, \Psi^{(1)} \right) \bigg) \\
  &+ Q\lambda_{2}^{(1,1)}\bigg( \Psi^{(1)}, Q^{-1}L_{2}\left( \Psi^{(1)}, \Psi^{(1)} \right) \bigg). \\
\end{split}
\end{equation}
Here and in the following, whenever the BRST charge $Q$ hits the propagator $Q^{-1}$ we use the chain homotopy relation
\begin{equation} 
1 - P_{0} = Q Q^{-1} + Q^{-1} Q
\end{equation}
until it finally hits an on-shell state and vanishes.
To continue we use the $L_{\infty}$ relation $[\mathbf{Q}, \mathbf{L}_2]=0$.
Furthermore, $P_{0} L_{2}\left( \Psi^{(1)}, \Psi^{(1)} \right) =0= P_{0} \lambda_{2}^{(1,1)}\left( \Psi^{(1)}, \Psi^{(1)} \right)$.
The first identity was already shown in subsection~\ref{subsec:secob}, while the second is due to
\begin{equation}\label{eq:vp0}
  \begin{split}
    P_{0} \lambda_{2}^{(1, 1)}(\Psi^{(1)}, \Psi^{(1)}) &= P_{0} \bar{\xi} \circ L_{2}^{(1, 0)} (\Psi^{(1)}, \Psi^{(1)}) \\
                                                       &= P_{0} (\mathcal{X}\circ\ell_{2})\left( cVe^{-\phi} , cVe^{-\phi} \right) \otimes P_{0}\bar{\xi} \circ  \ell_{2}\left( \bar{c}\bar{V}e^{-\bar{\phi}} , \bar{c}\bar{V}e^{-\bar{\phi}} \right ) \\
                                                       &= 0,
  \end{split}
\end{equation}
where $\bar{\xi} \circ L_{2}^{(1, 0)}$ is defined in appendix~\ref{app:sft}.
This brings~\eqref{eqn:obs2} into
\begin{equation}\label{eq:mass}
\begin{split}
  &-\lambda_{2}^{(1,1)}\bigg( \Psi^{(1)}, L_{2}\left( \Psi^{(1)}, \Psi^{(1)} \right) \bigg) + L_{2}\bigg( \Psi^{(1)}, \lambda_{2}^{(1,1)}\left( \Psi^{(1)}, \Psi^{(1)} \right) \bigg) + QL_{2}\bigg( \Psi^{(1)}, Q^{-1}\lambda_{2}^{(1,1)}\left( \Psi^{(1)}, \Psi^{(1)} \right) \bigg) \\
  &+ Q\lambda_{2}^{(1,1)}\bigg( \Psi^{(1)}, Q^{-1}L_{2}\left( \Psi^{(1)}, \Psi^{(1)} \right) \bigg). \\
\end{split}
\end{equation}
For the $L_3$ term in~\eqref{eq:o3} we use that by construction we have
\begin{equation}
\mathbf{L}_{3}^{(2, 2)} = \frac{1}{2} ([\mathbf{Q}, \boldsymbol{\lambda}_{3}^{(2, 2)}] + [\mathbf{L}_{2}^{(1, 1)}, \boldsymbol{\lambda}_{2}^{(1, 1)}]),
\end{equation}
canceling out the first two terms in \eqref{eq:mass}.
The obstruction reads
\begin{equation}
\begin{split}
  \mathcal{O}^{(3)}_i&=\left\langle e^{i}, -\frac{3}{2}\bar{\xi}QL_{2}\bigg( \Psi^{(1)}, Q^{-1}\lambda_{2}^{(1,1)}\left( \Psi^{(1)}, \Psi^{(1)} \right) \bigg) -\frac{3}{2} \bar{\xi}Q\lambda_{2}^{(1,1)}\bigg( \Psi^{(1)}, Q^{-1}L_{2}\left( \Psi^{(1)}, \Psi^{(1)} \right) \bigg) \right.\\
  &\left. + \frac{1}{2}\bar{\xi}Q\lambda_{3}^{(2,2)}\bigg( \Psi^{(1)}, \Psi^{(1)}, \Psi^{(1)} \bigg) \right\rangle_{S, L}\\
  &= \left\langle e^{i}, -\frac{3}{2}\bar{\mathcal{X}}L_{2}\bigg( \Psi^{(1)}, Q^{-1}\lambda_{2}^{(1,1)}\left( \Psi^{(1)}, \Psi^{(1)} \right) \bigg) - \frac{3}{2}\bar{\mathcal{X}}\lambda_{2}^{(1,1)}\bigg( \Psi^{(1)}, Q^{-1}L_{2}\left( \Psi^{(1)}, \Psi^{(1)} \right) \bigg) \right.\\
  &\left. + \frac{1}{2}\bar{\mathcal{X}}\lambda_{3}^{(2,2)}\bigg( \Psi^{(1)}, \Psi^{(1)}, \Psi^{(1)} \bigg) \right\rangle_{S, L},
\end{split}
\end{equation}
where the inner product is taken in the large Hilbert space for the anti-holomorphic sector only.
The $\bar{\xi}$ zero mode must be provided by $\lambda_2^{(1,1)}$ and $\lambda_3^{(2,2)}$, so
\begin{equation}
\begin{split}
  \mathcal{O}^{(3)}_i &= \left\langle e^{i}, -\frac{3}{2}\bar{\xi}\bar{\mathcal{X}}L_{2}\bigg( \Psi^{(1)}, Q^{-1}L_{2}^{(1,0)}\left( \Psi^{(1)}, \Psi^{(1)} \right) \bigg) - \frac{3}{2}\bar{\xi}\bar{\mathcal{X}}L_{2}^{(1,0)}\bigg( \Psi^{(1)}, Q^{-1}L_{2}\left( \Psi^{(1)}, \Psi^{(1)} \right) \bigg) \right.\\
  &\left. + \frac{1}{2}\bar{\xi}\bar{\mathcal{X}}L_{3}^{(2,1)}\bigg( \Psi^{(1)}, \Psi^{(1)}, \Psi^{(1)} \bigg) \right\rangle_{S, L} \\
  &= \left\langle e^{i}, -\frac{3}{2}\bar{\xi}\bar{\mathcal{X}}Q\lambda^{(1,1)}_{2}\bigg( \Psi^{(1)}, Q^{-1}L_{2}^{(1,0)}\left( \Psi^{(1)}, \Psi^{(1)} \right) \bigg) + \frac{3}{2}\bar{\xi}\bar{\mathcal{X}}\lambda_2^{(1,1)}\bigg( \Psi^{(1)}, QQ^{-1}L_{2}^{(1,0)}\left( \Psi^{(1)}, \Psi^{(1)} \right) \bigg) \right.\\
  &\left. - \frac{3}{2}\bar{\xi}\bar{\mathcal{X}}L_{2}^{(1,0)}\bigg( \Psi^{(1)}, Q^{-1}Q\lambda^{(1,1)}_{2}\left( \Psi^{(1)}, \Psi^{(1)} \right) \bigg) + \frac{1}{2}\bar{\xi}\bar{\mathcal{X}}L_{3}^{(2,1)}\bigg( \Psi^{(1)}, \Psi^{(1)}, \Psi^{(1)} \bigg) \right\rangle_{S, L},
\end{split}
\end{equation}
where we expand $L_{2}$ again using $\mathbf{L}_{2} = \left[ \mathbf{Q}, \boldsymbol{\lambda}_{2}^{(1,1)} \right]$.
Note that $Q$ is BPZ even, and hitting it with $\bar{\xi}$ gives $\bar{\mathcal{X}}$ since $Qe^{i}=0$.
We have, 
\begin{equation}
\begin{split}
  \mathcal{O}^{(3)}_i &= \left\langle e^{i}, -\frac{3}{2}\bar{\xi}\bar{\mathcal{X}}^{2}L^{(1,0)}_{2}\bigg( \Psi^{(1)}, Q^{-1}L_{2}^{(1,0)}\left( \Psi^{(1)}, \Psi^{(1)} \right) \bigg) \right.\\
  &\left. + \frac{3}{2}\bar{\xi}\bar{\mathcal{X}}\lambda_2^{(1,1)}\bigg( \Psi^{(1)}, \left( 1 - Q^{-1}Q - P_{0} \right)L_{2}^{(1,0)}\left( \Psi^{(1)}, \Psi^{(1)} \right) \bigg) \right.\\
  &\left. - \frac{3}{2}\bar{\xi}\bar{\mathcal{X}}L_{2}^{(1,0)}\bigg( \Psi^{(1)}, \left( 1 - QQ^{-1} - P_0 \right)\lambda^{(1,1)}_{2}\left( \Psi^{(1)}, \Psi^{(1)} \right) \bigg) \right.\\
  &\left. + \frac{1}{2}\bar{\xi}\bar{\mathcal{X}}L_{3}^{(2,1)}\bigg( \Psi^{(1)}, \Psi^{(1)}, \Psi^{(1)} \bigg) \right\rangle_{S, L} \\
  &= \left\langle e^{i}, -\frac{3}{2}\bar{\xi}\bar{\mathcal{X}}^{2}L^{(1,0)}_{2}\bigg( \Psi^{(1)}, Q^{-1}L_{2}^{(1,0)}\left( \Psi^{(1)}, \Psi^{(1)} \right) \bigg) \right.\\
  &\left. - \frac{3}{2}\bar{\xi}\bar{\mathcal{X}}L_{2}^{(1,0)}\bigg( \Psi^{(1)}, QQ^{-1}\lambda^{(1,1)}_{2}\left( \Psi^{(1)}, \Psi^{(1)} \right) \bigg) \right.\\
  &\left. + \frac{3}{2}\bar{\xi}\bar{\mathcal{X}}\lambda_2^{(1,1)}\bigg( \Psi^{(1)}, L_{2}^{(1,0)}\left( \Psi^{(1)}, \Psi^{(1)} \right) \bigg) \right.\\
  &\left. - \frac{3}{2}\bar{\xi}\bar{\mathcal{X}}L_{2}^{(1,0)}\bigg( \Psi^{(1)}, \lambda^{(1,1)}_{2}\left( \Psi^{(1)}, \Psi^{(1)} \right) \bigg) \right.\\
  &\left. + \frac{1}{2}\bar{\xi}\bar{\mathcal{X}}L_{3}^{(2,1)}\bigg( \Psi^{(1)}, \Psi^{(1)}, \Psi^{(1)} \bigg) \right\rangle_{S, L},
\end{split}
\end{equation}
where we used the fact that the terms involving $P_0$ vanish for similar reasons as in \eqref{eq:vp0}. 
Again, we see that the contact terms  cancel with the 3-product term by $\mathbf{L}_{3}^{(2, 1)} = \left[ \mathbf{Q}, \boldsymbol{\lambda}_{3}^{(2,1)} \right] + \left[ \mathbf{L}_2^{(1, 0)}, \boldsymbol{\lambda}_{2}^{(1,1)} \right]$, leaving
\begin{equation}
\begin{split}
  \mathcal{O}^{(3)}_i &= \left\langle e^{i}, -\frac{3}{2}\bar{\xi}\bar{\mathcal{X}}^{2}L^{(1,0)}_{2}\bigg( \Psi^{(1)}, Q^{-1}L_{2}^{(1,0)}\left( \Psi^{(1)}, \Psi^{(1)} \right) \bigg) - \frac{3}{2}\bar{\xi}\bar{\mathcal{X}}QL^{(1,0)}_{2}\bigg( \Psi^{(1)}, Q^{-1}\lambda_{2}^{(1,1)}\left( \Psi^{(1)}, \Psi^{(1)} \right) \bigg) \right.\\
  &\left. + \frac{1}{2}\bar{\xi}\bar{\mathcal{X}}Q\lambda_{3}^{(2,1)}\bigg( \Psi^{(1)}, \Psi^{(1)}, \Psi^{(1)} \bigg) \right\rangle_{S, L},
\end{split}
\end{equation}
so we can pull out the $\bar{\xi}$ zero mode again, and finally we arrive at
\begin{equation}
  \mathcal{O}^{(3)}_i = \left\langle \bar{\xi}\bar{\mathcal{X}}^{2}e^{i}, -3L^{(1,0)}_{2}\bigg( \Psi^{(1)}, Q^{-1}L_{2}^{(1,0)}\left( \Psi^{(1)}, \Psi^{(1)} \right) \bigg) + L_{3}^{(2,0)}\bigg( \Psi^{(1)}, \Psi^{(1)}, \Psi^{(1)} \bigg) \right\rangle_{S, L}.
\end{equation}
Next, we show that $\bar{\mathcal{X}}$ can be moved around, up to vanishing $P_{0}$ boundary terms and additional contact terms\footnote{We thank the anonymous reviewer for pointing this out.}.
Noting that upon taking inner product in the large Hilbert space, the $\bar{\xi}$ zero mode can be attached to an arbitrary vertex up to a minus sign, as long as the other parts are all in the small Hilbert space, we have
\begin{equation}\label{eqn:split}
  \begin{split}
    \mathcal{O}_{i}^{(3)} &= \left\langle Q\bar{\xi}\bar{\mathcal{X}}e^{i}, 3L_{2}^{(1,0)}\bigg( \bar{\xi}\Psi^{(1)}, Q^{-1}L_{2}^{(1,0)}\left( \Psi^{(1)}, \Psi^{(1)} \right) \bigg) - L_{3}^{(2,0)}\left(\bar{\xi}\Psi^{(1)}, \Psi^{(1)}, \Psi^{(1)}\right) \right\rangle_{S, L} \\
    &= \left\langle \bar{\xi}\bar{\mathcal{X}}e^{i}, 3QL_{2}^{(1,0)}\bigg( \bar{\xi}\Psi^{(1)}, Q^{-1}L_{2}^{(1,0)}\left( \Psi^{(1)}, \Psi^{(1)} \right) \bigg) - QL_{3}^{(2,0)}\left(\bar{\xi}\Psi^{(1)}, \Psi^{(1)}, \Psi^{(1)}\right) \right\rangle_{S, L} \\
                &= \left\langle \bar{\xi}\bar{\mathcal{X}}e^{i}, -3L_{2}^{(1,0)}\bigg( Q\bar{\xi}\Psi^{(1)}, Q^{-1}L_{2}^{(1,0)}\left( \Psi^{(1)}, \Psi^{(1)} \right) \bigg) \right.\\
                &\left. + 3L_{2}^{(1,0)}\bigg( \bar{\xi}\Psi^{(1)}, QQ^{-1}L_{2}^{(1,0)}\left( \Psi^{(1)}, \Psi^{(1)} \right) \bigg) \right. \\
                &\left. - QL_{3}^{(2,0)}\left(\bar{\xi}\Psi^{(1)}, \Psi^{(1)}, \Psi^{(1)}\right) \right\rangle_{S, L} \\
                &= \left\langle \bar{\xi}\bar{\mathcal{X}}e^{i}, -3L_{2}^{(1,0)}\bigg( \bar{\mathcal{X}}\Psi^{(1)}, Q^{-1}L_{2}^{(1,0)}\left( \Psi^{(1)}, \Psi^{(1)} \right) \bigg) \right.\\
                &\left. + 3L_{2}^{(1,0)}\bigg( \bar{\xi}\Psi^{(1)}, \left( 1-Q^{-1}Q-P_0 \right)L_{2}^{(1,0)}\left( \Psi^{(1)}, \Psi^{(1)} \right) \bigg) \right. \\
                &\left. - QL_{3}^{(2,0)}\left(\bar{\xi}\Psi^{(1)}, \Psi^{(1)}, \Psi^{(1)}\right) \right\rangle_{S, L} \\
                &= \left\langle \bar{\xi}\bar{\mathcal{X}}e^{i}, -3L_{2}^{(1,0)}\bigg( \bar{\mathcal{X}}\Psi^{(1)}, Q^{-1}L_{2}^{(1,0)}\left( \Psi^{(1)}, \Psi^{(1)} \right) \bigg) + L_{3}^{(2,0)}\left(\bar{\mathcal{X}}\Psi^{(1)}, \Psi^{(1)}, \Psi^{(1)}\right) \right\rangle_{S, L} \\
                & +\left\langle \bar{\xi}\bar{\mathcal{X}}e^{i}, 2L_{2}^{(1,0)}\bigg( \bar{\xi}\Psi^{(1)}, L_{2}^{(1,0)}\left( \Psi^{(1)}, \Psi^{(1)} \right) \bigg) \right.\\
                &\left. + 2L_{2}^{(1,0)}\bigg( \Psi^{(1)}, L_{2}^{(1,0)}\left( \bar{\xi}\Psi^{(1)}, \Psi^{(1)} \right) \bigg) \right\rangle_{S, L}.
  \end{split}
\end{equation}
For now let us focus on the first term in~\eqref{eqn:split}, denoted by $\mathcal{O}^{(3)}_{i, A}$, whereas the second term is denoted by $\mathcal{O}^{(3)}_{i, B}$.
We use the split $\Psi^{(1)} = \Psi^{(1)}_{+} + \Psi^{(1)}_{-}$, giving
\begin{equation}\label{eqn:first-contribution}
    \mathcal{O}_{i,A}^{(3)} = \sum_{a, b, c\in \{\pm\}}\left\langle \bar{\xi}\bar{\mathcal{X}}e^{i}, -3L_{2}^{(1,0)}\bigg( \bar{\mathcal{X}}\Psi^{(1)}_{a}, Q^{-1}L_{2}^{(1,0)}\left( \Psi^{(1)}_{b}, \Psi^{(1)}_{c} \right) \bigg) + L_{3}^{(2,0)}\left(\bar{\mathcal{X}}\Psi^{(1)}_{a}, \Psi^{(1)}_{b}, \Psi^{(1)}_{c}\right) \right\rangle_{S, L}.
\end{equation}
The remaining $U(1)$-charge decomposition and the explicit cancellation between $\mathcal{O}^{(3)}_{i,A}$ and $\mathcal{O}^{(3)}_{i,B}$ are presented in appendix~\ref{app:third-order}, where we show that $\mathcal{O}^{(3)}_i=0$.

To conclude this section, we have the result that there is no cohomological obstruction to resolving the orbifold to the third order, no matter what moduli we choose.
This result contrasts with the result in a similar open superstring setup~\cite{Mattiello2019} (and more generally~\cite{Vošmera2019}), where the ADHM constraint or its generalization needs to be imposed on the moduli, which is indeed a difference between the construction of the gauge and gravitational instantons.
This can be understood from the string theory perspective, since the factorization into holomorphic and anti-holomorphic parts guarantees that there is always a part vanishing in the obstruction.

\section{Conclusion and outlook}\label{sec:con}

This work studied the instantonic background of Type IIB string theory.
In particular, we focused on the resolved $\mathbb{C}^{2}/\mathbb{Z}_{2}$ orbifold, namely the Eguchi-Hanson space, which is one of the ALE classical gravitational instantons.
In this paper we recovered the Eguchi-Hanson metric from string field theory by tuning the moduli and taking the field theory limit.
$\alpha'$ corrections to the latter are explicit in eqn.~\eqref{eqn:metric}.
Moreover, we showed that a general deformation is hyperk\"ahler at this order.
At higher order, the computation becomes increasingly involved.
We showed the third order solution is not obstructed.

It remains an important task to establish an all-order background in closed string field theory.
In one of the simplest cases, the pp-wave background, an all-order solution has been considered~\cite{Cho2025} using an argument involving a worldsheet conserved charge.

One interesting phenomenon is that the holomorphic factorization in Type II theory drastically simplifies the derivation of the vanishing of the obstruction.
It therefore motivates the study of such deformations for heterotic theory, which is indeed more suitable for realistic models.

Another somewhat mysterious aspect is the K\"ahlerness of our solution, which implies an enhanced $\mathcal{N} = 2$ worldsheet supersymmetry.
In spacetime physics it is reflected as self-dual physics~\cite{Ooguri1990} and is closely related to integrability.
Consideration in the $\mathcal{N}=2$ string field theory~\cite{Zhang2025} framework may simplify the structure.

Finally, our results for a general choice of deformation moduli suggest a natural connection to generalized K\"ahler geometry.
While we recovered the standard hyperk\"ahler Eguchi-Hanson metric by specifically tuning the moduli such that the Kalb-Ramond field vanishes ($B_{ij}=0$), a generic marginal deformation naturally turns on a non-vanishing $B$-field.
In the presence of such an anti-symmetric tensor field, the target space geometry of the $\mathcal{N}=(2,2)$ worldsheet supersymmetric sigma model is known to be generalized K\"ahler.
Geometrically, the $B$-field can be interpreted as a connection on a bundle gerbe over the resolved orbifold.
Closed string field theory thus provides a systematic, order-by-order framework to explicitly construct these generalized K\"ahler instantons and their associated gerbe structures directly from the off-shell worldsheet CFT~\cite{Gates1984}.
It would be amusing to further explore this explicit construction.
We hope to come back to these in the future.

\acknowledgments
The work of I.S. is supported by the Excellence Cluster Origins of the DFG under Germany's Excellence Strategy EXC-2094 390783311 as well as EXC 2094/2: ORIGINS 2.
The work of X.-H.Z. is financially supported in part by JST SPRING, Grant Number JPMJSP2125. X.-H.Z. would like to thank the ``THERS Make New Standards Program for the Next Generation Researchers'' and ``2025 the 3rd
Overseas Travel Expense Support by MNS program.'' We thank the anonymous reviewer for pointing out a conceptual mistake regarding the obstruction in the first version of this manuscript.

\appendix
\section{Obstruction in bosonic theory}\label{app:obs}
In this appendix we show that the perturbative approach in closed bosonic string theory is obstructed already at the second order, motivating our consideration in a superstring setup.
As a preliminary step, we identify the massless fields and, in particular, the moduli of a $\mathbb{C}^{n}/\mathbb{Z}_{2}$ conifold in bosonic string theory.
Recalling that the ground state energy of a worldsheet boson/fermion with mode number $n + \theta$ is
\begin{equation}
E_{0} = \pm \frac{1}{48} \mp \frac{1}{16}(2\theta - 1)^{2}.
\end{equation}
We have in the untwisted sector $(\theta = 0)$
\begin{equation}
E_{0} = -(D - 2)\frac{1}{24} + \frac{c}{24} - 1,
\end{equation}
where $c$ is the central charge and the ghosts contribute a $-1$.
 The massless fields in the untwisted sector are given by
\begin{equation}
    \alpha_{-1}^{\mu} \tilde{\alpha}^{\nu}_{-1} |0 \rangle \quad \mathrm{and} \quad
    \left\{
    \begin{array}{c}
    \alpha_{-1}^{i} \tilde{\alpha}_{-1}^{\bar{j}} |0\rangle\\
    \alpha_{-1}^{\bar{i}} \tilde{\alpha}_{-1}^{j} |0\rangle
    \end{array}
\right.,
\end{equation}
where $\mu = 2, \cdots, 9$ and $\alpha_{-1}^{j} = \frac{1}{\sqrt{2}} \left( \alpha^{2k}_{-1} + i\alpha^{2k+1}_{-1}\right),$ $ \alpha_{-1}^{\bar{j}} = \frac{1}{\sqrt{2}} \left( \alpha^{2k}_{-1} - i\alpha^{2k+1}_{-1}\right),$ $ k = 2j$, $j = 5, \cdots, 12$ and similarly for $\tilde{\alpha}^{j}_{-1}$.

In the twisted sector with $\theta = \frac{1}{2}$ we have instead,
\begin{equation}
E_{0} = \frac{n}{16} - 1,
\end{equation}
where $n$ is the dimension of the orbifold and $-1$ is again the ghost contribution. The massless field in the twisted sector is then
\begin{equation}
V \equiv |0\rangle_{\theta} = \prod_{k=10}^{25} \sigma_{k} |0\rangle \equiv \Delta|0\rangle,
\end{equation}
where $\sigma_{k}$ is the $\mathbb{Z}_{2}$ twist field for $X^{k}$. $V$ together with the right moving partner $\tilde{V}$ is then the marginal field corresponding to the blow-up of the orbifold.
More precisely, $\psi_{0} = \tilde{c} \tilde{V} c V \in \mathcal{H}$ is tangent to the moduli space at the orbifold point.
The Hilbert space $\mathcal{H}$ is the direct sum of the two sectors,
\begin{equation}
\mathcal{H} = \mathcal{H}_{0} \oplus \mathcal{H}_{\theta}.
\end{equation}

The cubic interaction of bosonic string field theory gives rise to the $2$-product
\begin{equation}
    \ell_{2} : \mathcal{H} \otimes \mathcal{H} \to \mathcal{H}.
\end{equation}
In particular, $\mathcal{H}_{\theta} \otimes \mathcal{H}_{\theta}$ gets mapped into $\mathcal{H}_{0}$. In order to test exact marginality of $V$ we consider the projection

\begin{equation}
P_{0} \ell(\tilde{c}\tilde{V}cV, \tilde{c}\tilde{V}cV) \subset \lim_{\bar{z}\to0} P_{0} \left[ \tilde{c}\tilde{V}(\bar{z}) \tilde{c}\tilde{V}(-\bar{z}) \right] \otimes \lim_{z\to0} P_{0} \left[ cV(z) cV(-z) \right],
\end{equation}
where $P_{0} \subset \mathrm{ker}(L_{0}^{+}) \subset \mathrm{ker}(L_{0}) \cap \mathrm{ker}(\bar{L}_{0})$, is the projector on the cohomology of $Q$.
It has been argued (see~\cite{Mattiello2018}) that
\begin{equation}
\lim_{z\to0} P_{\mathrm{ker}(L_{0})} \left[ cV(z) cV(-z) \right] \sim c\partial c \left( e^{i\sqrt{2}X} - e^{-i\sqrt{2}X} \right)(0),
\end{equation}
which then implies an obstruction to exact marginality.

\section{Hyperk\"ahler geometry and ALE instanton}\label{app:geo}

In this appendix, we first review the basics of hyperkähler geometry and asymptotically locally Euclidean (ALE) gravitational instantons.
We also prove that the spacetime geometry $W_{ij}^{(2)}$ that we derived in subsection~\ref{subsec:gra} is hyperkähler to the second order.

\subsection{Hyperkähler Manifolds}

A \textbf{gravitational instanton} is defined as a complete, non-compact, Riemannian 4-manifold $(M, g)$ that is Ricci-flat
\begin{equation}
R(g) = 0,
\end{equation}
and has curvature vanishing at infinity.
A Riemannian manifold $(M, g)$ is \textbf{Kähler} if it admits a complex structure $J: TM \to TM, J^{2} = -1$ that is compatible with the metric and is parallel with respect to the Levi-Civita connection
\begin{equation}
\nabla_{g} J = 0,
\end{equation}
where $\nabla_{g}$ is the covariant derivative,
such that the fundamental 2-form $\omega$ defined as $\omega(\cdot, \cdot) = g(J\cdot, \cdot)$ is closed, i.e. $\mathrm{d}\omega = 0$.

A \textbf{hyperkähler manifold} is a special type of Riemannian manifold endowed with three distinct complex structures $(I, J, K)$ that are compatible with the metric and satisfy the quaternion algebra relations:
\begin{equation}
I^{2} = J^{2} = K^{2} = IJK = -1.
\end{equation}
Again, we have $\nabla I = \nabla J = \nabla K = 0$.
Consequently, we can define three distinct Kähler forms:
\begin{equation}
\omega_I(\cdot, \cdot) = g(I\cdot, \cdot), \quad \omega_J(\cdot, \cdot) = g(J\cdot, \cdot), \quad \omega_K(\cdot, \cdot) = g(K\cdot, \cdot).
\end{equation}
A 4-dimensional manifold with a hyperkähler structure must have a holonomy group contained in $Sp(1) \cong SU(2) \subset SO(4)$.
Since $SU(2)$ is the holonomy group of a Calabi-Yau 2-fold, all hyperkähler 4-manifolds are necessarily Ricci-flat.
Thus, a hyperkähler perturbation to the Euclidean space is automatically a gravitational instanton.


\subsection{ALE Spaces and Resolutions}

Gravitational instantons are classified by their asymptotic behavior.
Here we are interested in the \textbf{asymptotically locally Euclidean (ALE)} spaces with quotient flat metric at infinity.
We take $\Gamma \subset SU(2)$ to be a finite discrete subgroup acting freely on $\mathbb{R}^4 \setminus \{0\}$.
For the classical ADE classification, if $\Gamma = \mathbb{Z}_k$, the resulting space is of type $A_{k-1}$.

Our main text focuses on the $\mathbb{Z}_2$ orbifold that is of Type $A_1$.
The quotient $\mathbb{C}^2 / \mathbb{Z}_2$ has an orbifold singularity at the origin.
The gravitational instanton corresponds to the resolution of this singularity.
Topologically, the resolution replaces the singular point with a collection of 2-spheres (exceptional divisors) $E_i$ having self-intersection number $-2$.
For $\mathbb{Z}_2$, there is a single exceptional 2-sphere $\mathbb{C}P^1$, and the manifold is diffeomorphic to the cotangent bundle $T^* \mathbb{C}P^1$.
The metric on the resolution of $\mathbb{C}^2/\mathbb{Z}_2$ is the celebrated \textbf{Eguchi-Hanson metric}.
It can be constructed explicitly via a Kähler potential as we show in the following.

Let $z \in \mathbb{C}^2$ be standard coordinates covering the space (away from the zero section).
The potential is given by:
\begin{equation}
\mathcal{K}_{EH}(z) = \frac{1}{2}\left( \sqrt{1 + |z|^4} + 2\log|z| - \log(1 + \sqrt{1+|z|^4}) \right).
\end{equation}
The associated Kähler form $\omega_{EH} = i\partial\bar{\partial}\mathcal{K}_{EH}$ yields the metric
\begin{equation}
  H_{I\bar{J}} = {(1+\frac{\rho^{4}}{r^{4}})}^{1/2}
  \begin{pmatrix} 1-\frac{\rho^{4} \bar{z}_{1} z_{1}}{r^2(\rho^4+r^4)} & -\frac{\rho^4 \bar{z}_{1}z_{2}}{r^2(\rho^4+r^4)} \\
    -\frac{\rho^4 \bar{z}_{2} z_{1}}{r^2(\rho^4+r^4)} & 1 -\frac{\rho^4 \bar{z}_{2} z_{2}}{r^2(\rho^4+r^4)} \end{pmatrix},
\end{equation}
where $r^2 = |z|^2$ and $\rho$ is the size of the blown-up 2-sphere.
As $|z| \to \infty$, this potential approaches $\mathcal{K}_{flat} \sim |z|^2/2$, recovering the flat metric on the orbifold $\mathbb{C}^2/\mathbb{Z}_2$.
In the limit $\rho \to 0$, we recover the singular orbifold metric.

\subsection{General metric profile}\label{app:weyl}
A general choice of moduli gives
\begin{equation}
W_{ij}^{(2)} = - \frac{4\mathcal{N}\left( \alpha' ,k\right)}{k^{2}} \mathbf{w}_{i} \bar{\mathbf{w}}_{j}
\end{equation}
where the vector $\mathbf{w}$ is defined as:
\begin{equation}
\mathbf{w} =
\begin{pmatrix}
-(2 k_2 u + k_4 v) \\
2 k_1 u - k_3 v \\
-(2 k_4 u - k_2 v) \\
2 k_3 u + k_1 v
\end{pmatrix},\quad
\bar{\mathbf{w}} =
\begin{pmatrix}
  -(2 k_2 \bar{u} + k_4 \bar{v}) \\
  2 k_1 \bar{u} - k_3 \bar{v} \\
  -(2 k_4 \bar{u} - k_2 \bar{v}) \\
  2 k_3 \bar{u} + k_1 \bar{v}
\end{pmatrix},
\end{equation}
where $u = w_{1}w_{2}$, $v = w_{1}^{2} - w_{2}^{2}$, $\bar{u} = \bar{w}_{1}\bar{w}_{2}$, $\bar{v} = \bar{w}_{1}^{2} - \bar{w}_{2}^{2}$, $\mathcal{N}\left( \alpha' \right) = \left| f_{1}'(0) \right|^{\alpha' k^{2}}/16$.

\begin{equation}
  \Phi^{(2)} = \frac{1}{4} \mathrm{Tr} \left( W_{ij}^{(2)} \right) = - 4\mathcal{N}(\alpha') \left( 4u\bar{u} + v\bar{v} \right)
\end{equation}
The total profile in the field theory limit $\alpha'\to 0$ reads   
\begin{equation}\label{eqn:profile}
  \begin{split}
  (G_{ij}^{(2)}+B_{ij}^{(2)})&\left( x; u,v,\bar{u},\bar{v} \right) = -\frac{1}{x^{6}} \\
    &\times
\begin{pmatrix}
 -2 u \bar{u} \left(x^2-4 x_{2}^2\right) & -2 (2 u x_{2}+v x_{4}) & u \bar{v} \left(x^2-4 x_{2}^2\right) & -2 (2 u x_{2}+v x_{4}) \\
-\frac{1}{2} v \bar{v} \left(x^2-4 x_{4}^2\right) & \times (2 \bar{u} x_{1}-\bar{v} x_{3}) & -\bar{u} v \left(x^2-4 x_{4}^2\right) & \times (2 \bar{u} x_{3}+\bar{v} x_{1}) \\
+4 \left( u \bar{v} + \bar{u} v \right) x_{2} x_{4} &&+2 \left( 4u \bar{u} - v \bar{v} \right) x_{2} x_{4}& \\
  \\
 -2 (2 u x_{1}-v x_{3}) & -2 u\bar{u} \left(x^2-4 x_{1}^2\right) & 2 (2 u x_{1}-v x_{3}) & -u \bar{v} \left(x^2-4 x_{1}^2\right) \\
 \times (2 \bar{u} x_{2}+\bar{v} x_{4}) & -\frac{1}{2} v\bar{v} \left(x^2-4 x_{3}^2\right) & \times (\bar{v} x_{2}-2 \bar{u} x_{4}) & +\bar{u} v \left(x^2-4 x_{3}^2\right) \\
&-4 \left( u\bar{v}+\bar{u}v \right) x_{1} x_{3}&&+2 \left( 4 u \bar{u} - v \bar{v} \right)x_{1} x_{3} \\
  \\
 -u \bar{v} \left(x^2-4 x_{4}^2\right) & 2 (v x_{2}-2 u x_{4}) &  - 2 u\bar{u} \left(x^2-4 x_{4}^2\right) & 2 (v x_{2}-2 u x_{4}) \\
+\bar{u} v \left(x^2-4 x_{2}^2\right) & \times (2 \bar{u} x_{1}-\bar{v} x_{3}) & -\frac{1}{2} v\bar{v} \left(x^2-4 x_{2}^2\right) & \times (2 \bar{u} x_{3}+\bar{v} x_{1}) \\
+2 \left( 4 u \bar{u} - v \bar{v} \right) x_{2} x_{4} &&-4 \left( u\bar{v} + \bar{u}v \right) x_{2} x_{4}&\\
  \\
 -2 (2 u x_{3}+v x_{1}) & +u \bar{v} \left(x^2-4 x_{3}^2\right) & 2 (2 u x_{3}+v x_{1}) & -2 u \bar{u} \left(x^2-4 x_{3}^2\right) \\
 \times (2 \bar{u} x_{2}+\bar{v} x_{4}) & -\bar{u} v \left(x^2-4 x_{1}^2\right) & \times (\bar{v} x_{2}-2 \bar{u} x_{4}) & -\frac{1}{2} v \bar{v} \left(x^2-4 x_{1}^2\right) \\
  &+2 \left( 4u \bar{u} - v \bar{v} \right) x_{1} x_{3}&&+4 \left( u \bar{v} + \bar{u} v \right) x_{1} x_{3}
\end{pmatrix}
  \end{split}.
\end{equation}
An explicit calculation shows  that $G_{ij}^{(2)}(x)$ is Ricci flat to this order.
Furthermore, the linearized Weyl tensor obtained from $G_{ij}^{(2)}(x)$ is then (see e.g.~\cite{Wald1984})
\begin{equation}
  \begin{split}
  C_{ijkl} &= R_{ijkl} - \delta_{i[k}R_{l]j} + \frac{1}{3} R \delta_{i[k}\delta_{l]j} \\
    &= R_{ijkl}
  \end{split}.
\end{equation}
We can decompose it into $C^{+}$, $C^{-}$ denoting the self-dual part and anti-self-dual part under Hodge dual operation, respectively.
By explicit evaluation we get
\begin{equation}
  \begin{split}
C^{+}_{ijkl} &= \frac{1}{2} \left[ C_{ijkl} + \frac{1}{2} \sqrt{|g|} g^{mp} g^{nq} \epsilon_{ijmn} C_{pqkl} \right] \\
             &= \frac{1}{2} \left[ C_{ijkl} + \frac{1}{2} \epsilon_{ij}{}^{pq} C_{klpq} \right] \\
    &= C_{ijkl},
  \end{split}
\end{equation}
where $\epsilon_{ijkl}$ is the Levi-Civita symbol.  Thus the Weyl tensor of $G_{ij}^{(2)}(x)$ is self-dual away from the origin.

\section{Closed string products}\label{app:sft}

In this appendix, we review the explicit construction of the closed superstring products.
We begin with the bosonic string products, which are the initial data for the superstring recursion (see~\cite{Erler2020} for a comprehensive introduction).

The superstring vertices are constructed recursively from the products of \textit{bosonic} closed string field theory, denoted $l_{n}$ and also $L_n^{(0,0)}$, when interpreted as the initial data for the recursion.
These products are defined geometrically via the decomposition of the moduli space of Riemann surfaces.

Let $\mathcal{M}_{0,n}$ be the moduli space of genus zero Riemann surfaces with $n$ punctures.
Closed string field theory requires a decomposition of $\mathcal{M}_{0,n}$ into vertex regions $\mathcal{V}_{0,n}$ and propagator regions.
The $n$-string product $l_{n}$ is associated with the vertex region $\mathcal{V}_{0,n+1}$.

For a set of closed string states $\Phi_1, \dots, \Phi_n$, the product is defined by a surface state $\langle \Sigma_{0,n+1} |: \mathcal{H}^{\otimes (n+1)} \to \mathbb{C}$ associated with the region $\mathcal{V}_{0,n+1}$.
The defining relation is:
\begin{equation}
    \omega\big(\Phi, l_n(\Phi_1, \dots, \Phi_n)\big) = \int_{\mathcal{V}_{0,n+1}} \Omega_{n+1}(\Phi, \Phi_1, \dots, \Phi_n).
\end{equation}
Here, $\Omega_{n+1}$ is a volume form on the moduli space constructed from the string fields and the $b$-ghost insertions required to saturate the integration measure.
Explicitly, if $\tau^{\alpha}$ are local coordinates on $\mathcal{M}_{0,n+1} $, the measure is given by the insertion of the $b$-ghost field associated with the vector fields corresponding to the deformations of these coordinates:
\begin{equation}
    \mathcal{B} = \sum_{\alpha} b(\partial_{\alpha}) d\tau^\alpha.
\end{equation}
The bosonic product can be written formally as an integral over the vertex region of the moduli space:
\begin{equation}
    l_n(\Phi_1, \dots, \Phi_n) = \int_{\mathcal{V}_{0,n+1}} d\mu \, \left( \sum_{\sigma \in S_n} \mathcal{B} \cdot | \Phi_{\sigma(1)} \dots \Phi_{\sigma(n)} \rangle \right),
\end{equation}
where the states are mapped to the punctures of the sphere via conformal maps inherent to the definition of $\mathcal{V}_{0,n+1}$.
For $n=2$, there is a choice called the Witten vertex defined by
\begin{equation}
  f_{1}(z_{1}) = e^{2\pi i/3}\left( \frac{1 + iz_{1}}{1 - iz_{1}} \right)^{2/3}, \quad f_{2}(z_{2}) = \left( \frac{1 + iz_{2}}{1 - iz_{2}} \right)^{2/3}, \quad f_{3}(z_{3}) = e^{-2\pi i/3}\left( \frac{1 + iz_{3}}{1 - iz_{3}} \right)^{2/3}.
\end{equation}
For $n \ge 3$, these products should cover the regions of moduli space not generated by Feynman diagrams of lower-order vertices and the propagator, but we do not need the construction of them in this paper.

For the NS-NS sector of Type II superstring field theory, the vertices must carry picture number $(-1, -1)$, which means that the $n$-string product $L_{n}$ must carry total picture number $(n - 1, n - 1)$.
We illustrate the asymmetric construction in~\cite{Erler2014}, which builds the superstring products from the picture-zero bosonic products $L_{n}^{(0,0)}$.
Here the superscripts denote the left- and right-moving picture numbers, respectively, and thus the superstring products we need are denoted by $L_{n}^{(n-1, n-1)} \equiv L_{n}$.
The construction is a two-step procedure:
First, we construct products $L_{n+1}^{(n,0)}$ possessing the correct left-moving picture but zero right-moving picture.
We introduce a sequence of gauge products $\lambda^{(k,0)}$ constructed using the left-moving picture changing zero mode $\xi_0$.
The recursion proceeds by climbing a ``ladder'' of products: $L_{n+1}^{(0,0)}, \dots, L_{n+1}^{(n,0)}$, which is defined recursively
\begin{equation}
    \mathbf{L}_{m+n+2}^{(m+1, 0)} = \frac{1}{m+1} \sum_{k=0}^{m} \sum_{l=0}^{n} [\mathbf{L}_{k+l+1}^{(k, 0)}, \boldsymbol{\lambda}_{m+n-k-l+2}^{(m-k+1, 0)}],
\end{equation}
where the gauge products $\lambda_{n}^{(m,0)}$ are defined explicitly by:
\begin{equation}
    \lambda_{N+2}^{(m+1, 0)} = \frac{N-m+1}{N+3} \left( \xi_0 L_{N+2}^{(m, 0)} - L_{N+2}^{(m, 0)} (\xi_0 \wedge \mathbb{I}_{N+1}) \right)
.
\end{equation}
In the second step, we take the result $L_{n+1}^{(n,0)}$ from the first ladder and apply a similar recursion to the right-moving sector using the right-moving zero mode $\bar{\xi}_0$.
We climb the second ladder $L_{n+1}^{(n,0)}, \dots, L_{n+1}^{(n,n)}$ using the gauge products $\lambda^{(n, k)}$ defined by $\bar{\xi}_0$.
The recursive relations are similar to those in the first step but act on the right-moving sector:
\begin{equation}
    \mathbf{L}_{m+n+2}^{(m+n+1, m+1)} = \frac{1}{m+1} \sum_{k=0}^{m} \sum_{l=0}^{n} [\mathbf{L}_{k+l+1}^{(k+l, k)}, \boldsymbol{\lambda}_{m+n-k-l+2}^{(m+n-k-l+1, m-k+1)}],
\end{equation}
\begin{equation}
    \lambda_{N+2}^{(n, k+1)} = \frac{N-k+1}{N+3} \left( \bar{\xi}_0 L_{N+2}^{(n, k)} - L_{N+2}^{(n, k)} (\bar{\xi}_0 \wedge \mathbb{I}_{N+1}) \right)\equiv \bar{\xi} \circ L_{N+2}^{(n,k)}.
\end{equation}
We end this appendix by listing all the defining relations that are needed in the main text:

\begin{align}
\mathbf{L}^{(1,1)}_{2} &= [\mathbf{Q}, \boldsymbol{\lambda}^{(1,1)}_{2}],\\
\lambda^{(1,1)}_{2} &= \bar{\xi} \circ L_{2}^{(1,0)},\\
\mathbf{L}^{(2,2)}_{3} &= \frac{1}{2} \left( [\mathbf{Q}, \boldsymbol{\lambda}^{(2,2)}_{3}] + [\mathbf{L}^{(1,1)}_{2}, \boldsymbol{\lambda}^{(1,1)}_{2}] \right),\\
\lambda^{(2,2)}_{3} &= \bar{\xi} \circ L_{3}^{(2,1)},\\
\mathbf{L}_{3}^{(2, 1)} &= [ \mathbf{Q}, \boldsymbol{\lambda}_{3}^{(2,1)} ] + [ \mathbf{L}_2^{(1, 0)}, \boldsymbol{\lambda}_{2}^{(1,1)} ],\\
\lambda^{(2,1)}_{3} &= 2\bar{\xi} \circ L_{3}^{(2,0)}.
\end{align}

\section{Details of the third-order obstruction}\label{app:third-order}

In this appendix, we complete the $U(1)$-charge decomposition of the two contributions defined in~\eqref{eqn:split}, starting from the expression for $\mathcal{O}^{(3)}_{i,A}$ in~\eqref{eqn:first-contribution}, and exhibit their explicit cancellation.
The picture changed vertex is
\begin{equation}
\bar{\mathcal{X}}\Psi^{(1)}_a = \bar c\Phi_{0} + \bar \gamma \Phi_{a},
\end{equation}
with $\Phi_0$ neutral and $\Phi_{a}$ having the same $U(1)$ charge as $\Psi^{(1)}_{a}$.
The $\bar b\bar c$ ghost number conservation selects only the $\Phi_{0}$ contribution, and hence the remaining two vertices must have $U(1)$ charges that add up to zero.
We can then pick up the $U(1)$ charge obeying terms, at the price of producing more contact terms,
\begin{equation}
\begin{split}
                \mathcal{O}_{i,A}^{(3)} &= \sum_{\substack{a, b, c\in \{\pm\}\\b+c=0}}\left\langle \bar{\mathcal{X}}e^{i}, 3L_{2}^{(1,0)}\bigg( \bar{\mathcal{X}}\Psi^{(1)}_{a}, Q^{-1}L_{2}^{(1,0)}\left( \bar{\xi}\Psi^{(1)}_{b}, \Psi^{(1)}_{c} \right) \bigg) - L_{3}^{(2,0)}\left(\bar{\mathcal{X}}\Psi^{(1)}_{a}, \bar{\xi}\Psi^{(1)}_{b}, \Psi^{(1)}_{c}\right) \right\rangle_{S, L} \\
                &= \sum_{\substack{a, b, c\in \{\pm\}\\b+c=0\\a+c=0}}\left\langle \bar{\mathcal{X}}e^{i}, 3L_{2}^{(1,0)}\bigg( \Psi^{(1)}_{a}, Q^{-1}L_{2}^{(1,0)}\left( \bar{\mathcal{X}}\Psi^{(1)}_{b}, \bar{\xi}\Psi^{(1)}_{c} \right) \bigg) - L_{3}^{(2,0)}\left(\Psi^{(1)}_{a}, \bar{\mathcal{X}}\Psi^{(1)}_{b}, \bar{\xi}\Psi^{(1)}_{c}\right) \right\rangle_{S, L} \\
                              &+ \sum_{\substack{a, b, c\in \{\pm\}\\b+c=0}}\left\langle \bar{\mathcal{X}}e^{i}, 2L_{2}^{(1,0)}\bigg( \bar{\xi}\Psi^{(1)}_a, L_{2}^{(1,0)}\left( \bar{\xi}\Psi^{(1)}_b, \Psi^{(1)}_c \right) \bigg) + L_{2}^{(1,0)}\bigg(  L_{2}^{(1,0)}\left(\bar{\xi}\Psi^{(1)}_a, \bar{\xi}\Psi^{(1)}_b\right) , \Psi^{(1)}_c \bigg)\right. \\
                              &\left.+ L_{2}^{(1,0)}\bigg(  L_{2}^{(1,0)}\left(\Psi^{(1)}_c, \bar{\xi}\Psi^{(1)}_a\right) , \bar{\xi}\Psi^{(1)}_b \bigg) \right\rangle_{S, L} \\
                &= \sum_{\substack{a, b, c\in \{\pm\}\\b+c=0\\a+c=0\\a+b=0}}\left\langle \bar{\mathcal{X}}e^{i}, 3L_{2}^{(1,0)}\bigg( \Psi^{(1)}_{a}, Q^{-1}L_{2}^{(1,0)}\left( \bar{\xi}\Psi^{(1)}_{b}, \bar{\mathcal{X}}\Psi^{(1)}_{c} \right) \bigg) - L_{3}^{(2,0)}\left(\Psi^{(1)}_{a}, \bar{\xi}\Psi^{(1)}_{b}, \bar{\mathcal{X}}\Psi^{(1)}_{c}\right) \right\rangle_{S, L} \\
                              &+ \sum_{\substack{a, b, c\in \{\pm\}\\b+c=0\\a+c=0}}\left\langle \bar{\mathcal{X}}e^{i}, L_{2}^{(1,0)}\bigg( L_{2}^{(1,0)}\left( \Psi^{(1)}_a, \bar{\xi}\Psi^{(1)}_b, \right) \bar{\xi}\Psi^{(1)}_c \bigg) - L_{2}^{(1,0)}\bigg(  L_{2}^{(1,0)}\left(\Psi^{(1)}_a, \bar{\xi}\Psi^{(1)}_c\right) , \bar{\xi}\Psi^{(1)}_b \bigg)\right. \\
                              &\left.- 2L_{2}^{(1,0)}\bigg(  L_{2}^{(1,0)}\left(\bar{\xi}\Psi^{(1)}_b, \bar{\xi}\Psi^{(1)}_c\right) , \Psi^{(1)}_a \bigg) \right\rangle_{S, L} \\
                              &+ \sum_{\substack{a, b, c\in \{\pm\}\\b+c=0}}\left\langle \bar{\mathcal{X}}e^{i}, 2L_{2}^{(1,0)}\bigg( \bar{\xi}\Psi^{(1)}_a, L_{2}^{(1,0)}\left( \bar{\xi}\Psi^{(1)}_b, \Psi^{(1)}_c \right) \bigg) + L_{2}^{(1,0)}\bigg(  L_{2}^{(1,0)}\left(\bar{\xi}\Psi^{(1)}_a, \bar{\xi}\Psi^{(1)}_b\right) , \Psi^{(1)}_c \bigg)\right. \\
                              &\left.+ L_{2}^{(1,0)}\bigg(  L_{2}^{(1,0)}\left(\Psi^{(1)}_c, \bar{\xi}\Psi^{(1)}_a\right) , \bar{\xi}\Psi^{(1)}_b \bigg) \right\rangle_{S, L}.
  \end{split}
\end{equation}
Now for terms with two $\bar{\xi}$ insertions, we can insert $1=\bar{\xi}\bar{\eta}+\bar{\eta}\bar{\xi}$ to bring one $\bar{\xi}$ to the front of $\bar{\mathcal{X}}e^i$, that is,
\begin{equation}
\begin{split}
  \mathcal{O}_{i,A}^{(3)} &= \sum_{\substack{a, b, c\in \{\pm\}\\b+c=0}}\left\langle \bar{\xi}\bar{\mathcal{X}}e^{i}, L_{2}^{(1,0)}\bigg( L_{2}^{(1,0)}\left( \Psi^{(1)}_a,  \bar{\xi}\Psi^{(1)}_b \right), \Psi^{(1)}_c \bigg) - L_{2}^{(1,0)}\bigg(  L_{2}^{(1,0)}\left(\bar{\xi}\Psi^{(1)}_a, \Psi^{(1)}_b\right) , \Psi^{(1)}_c \bigg)\right. \\
&+ L_{2}^{(1,0)}\bigg(  L_{2}^{(1,0)}\left(\Psi^{(1)}_c, \Psi^{(1)}_a\right) , \bar{\xi}\Psi^{(1)}_b \bigg) - L_{2}^{(1,0)}\bigg(  L_{2}^{(1,0)}\left(\Psi^{(1)}_c, \bar{\xi}\Psi^{(1)}_a\right) , \Psi^{(1)}_b \bigg) \\
&\left.+ 2L_{2}^{(1,0)}\bigg(  L_{2}^{(1,0)}\left(\Psi^{(1)}_c, \Psi^{(1)}_b\right) , \bar{\xi}\Psi^{(1)}_a \bigg) - 2L_{2}^{(1,0)}\bigg(  L_{2}^{(1,0)}\left(\Psi^{(1)}_c, \bar{\xi}\Psi^{(1)}_b\right) , \Psi^{(1)}_a \bigg) \right\rangle_{S, L} \\
&+\sum_{\substack{a, b, c\in \{\pm\}\\b+c=0\\a+c=0}} \left\langle \bar{\xi}\bar{\mathcal{X}}e^{i}, L_{2}^{(1,0)}\bigg( L_{2}^{(1,0)}\left( \Psi^{(1)}_a,  \Psi^{(1)}_b \right), \bar{\xi}\Psi^{(1)}_c \bigg) - L_{2}^{(1,0)}\bigg(  L_{2}^{(1,0)}\left(\Psi^{(1)}_a, \bar{\xi}\Psi^{(1)}_b\right) , \Psi^{(1)}_c \bigg)\right. \\
&- L_{2}^{(1,0)}\bigg(  L_{2}^{(1,0)}\left(\Psi^{(1)}_a, \Psi^{(1)}_c\right) , \bar{\xi}\Psi^{(1)}_b \bigg) + L_{2}^{(1,0)}\bigg(  L_{2}^{(1,0)}\left(\Psi^{(1)}_a, \bar{\xi}\Psi^{(1)}_c\right) , \Psi^{(1)}_b \bigg) \\
&\left.- 2L_{2}^{(1,0)}\bigg(  L_{2}^{(1,0)}\left(\Psi^{(1)}_b, \bar{\xi}\Psi^{(1)}_c\right) , \Psi^{(1)}_a \bigg) + 2L_{2}^{(1,0)}\bigg(  L_{2}^{(1,0)}\left(\bar{\xi}\Psi^{(1)}_b, \Psi^{(1)}_c\right) , \Psi^{(1)}_a \bigg) \right\rangle_{S, L}.
\end{split}
\end{equation}
Explicitly writing down the charge sum we get
\begin{equation}
\begin{split}
  \mathcal{O}_{i,A}^{(3)} &= 2\left\langle \bar{\xi}\bar{\mathcal{X}}e^{i}_-, -L_{2}^{(1,0)}\bigg( L_{2}^{(1,0)}\left( \Psi^{(1)}_+,  \bar{\xi}\Psi^{(1)}_- \right), \Psi^{(1)}_+ \bigg) - L_{2}^{(1,0)}\bigg(  L_{2}^{(1,0)}\left(\bar{\xi}\Psi^{(1)}_+, \Psi^{(1)}_-\right) , \Psi^{(1)}_+ \bigg)\right. \\
&+ L_{2}^{(1,0)}\bigg(  L_{2}^{(1,0)}\left(\Psi^{(1)}_+, \Psi^{(1)}_+\right) , \bar{\xi}\Psi^{(1)}_- \bigg) - L_{2}^{(1,0)}\bigg(  L_{2}^{(1,0)}\left(\Psi^{(1)}_+, \bar{\xi}\Psi^{(1)}_+\right) , \Psi^{(1)}_- \bigg) \\
&\left.+ 2L_{2}^{(1,0)}\bigg(  L_{2}^{(1,0)}\left(\Psi^{(1)}_-, \Psi^{(1)}_+\right) , \bar{\xi}\Psi^{(1)}_+ \bigg) \right\rangle_{S, L} \\
&+ 2\left\langle \bar{\xi}\bar{\mathcal{X}}e^{i}_+, -L_{2}^{(1,0)}\bigg( L_{2}^{(1,0)}\left( \Psi^{(1)}_-,  \bar{\xi}\Psi^{(1)}_+ \right), \Psi^{(1)}_- \bigg) - L_{2}^{(1,0)}\bigg(  L_{2}^{(1,0)}\bar{\xi}\left(\Psi^{(1)}_-, \Psi^{(1)}_+\right) , \Psi^{(1)}_- \bigg)\right. \\
&+ L_{2}^{(1,0)}\bigg(  L_{2}^{(1,0)}\left(\Psi^{(1)}_-, \Psi^{(1)}_-\right) , \bar{\xi}\Psi^{(1)}_+ \bigg) - L_{2}^{(1,0)}\bigg(  L_{2}^{(1,0)}\left(\Psi^{(1)}_-, \bar{\xi}\Psi^{(1)}_-\right) , \Psi^{(1)}_+ \bigg) \\
&\left.+ 2L_{2}^{(1,0)}\bigg(  L_{2}^{(1,0)}\left(\Psi^{(1)}_+, \Psi^{(1)}_-\right) , \bar{\xi}\Psi^{(1)}_- \bigg) \right\rangle_{S, L}.
\end{split}
\end{equation}
We have now reduced $\mathcal{O}_{i,A}^{(3)}$ into a form which can be compared with the corresponding R-charge decomposition of the contact terms $\mathcal{O}_{i,B}^{(3)}$ in~\eqref{eqn:split}.
Writing out all charge combinations in $\mathcal{O}_{i,B}^{(3)}$ straightforwardly gives
\begin{equation}
\begin{split}
\mathcal{O}_{i,B}^{(3)} &= \sum_{\substack{a,b,c,d\in{\pm}\\a+b+c+d=0}}\left\langle \bar{\xi}\bar{\mathcal{X}}e^{i}_a, 2L_{2}^{(1,0)}\bigg( \bar{\xi}\Psi^{(1)}_b, L_{2}^{(1,0)}\left( \Psi^{(1)}_c, \Psi^{(1)}_d \right) \bigg) + 2L_{2}^{(1,0)}\bigg( \Psi^{(1)}_b, L_{2}^{(1,0)}\left( \bar{\xi}\Psi^{(1)}_c, \Psi^{(1)}_d \right) \bigg) \right\rangle_{S, L} \\
&= \left\langle \bar{\xi}\bar{\mathcal{X}}e^{i}_+, -2L_{2}^{(1,0)}\bigg( L_{2}^{(1,0)}\left( \Psi^{(1)}_-, \Psi^{(1)}_- \right), \bar{\xi}\Psi^{(1)}_+ \bigg) + 2L_{2}^{(1,0)}\bigg( L_{2}^{(1,0)}\left( \bar{\xi}\Psi^{(1)}_-, \Psi^{(1)}_- \right), \Psi^{(1)}_+ \bigg) \right\rangle_{S, L} \\
&+ \left\langle \bar{\xi}\bar{\mathcal{X}}e^{i}_+, -4L_{2}^{(1,0)}\bigg( L_{2}^{(1,0)}\left( \Psi^{(1)}_+, \Psi^{(1)}_- \right), \bar{\xi}\Psi^{(1)}_- \bigg) + 2L_{2}^{(1,0)}\bigg( L_{2}^{(1,0)}\left( \bar{\xi}\Psi^{(1)}_+, \Psi^{(1)}_- \right), \Psi^{(1)}_- \bigg) \right\rangle_{S, L} \\
&+ \left\langle \bar{\xi}\bar{\mathcal{X}}e^{i}_+, 2L_{2}^{(1,0)}\bigg( L_{2}^{(1,0)}\left( \bar{\xi}\Psi^{(1)}_-, \Psi^{(1)}_+ \right), \Psi^{(1)}_- \bigg) \right\rangle_{S, L} \\
&+ \left\langle \bar{\xi}\bar{\mathcal{X}}e^{i}_-, -2L_{2}^{(1,0)}\bigg( L_{2}^{(1,0)}\left( \Psi^{(1)}_+, \Psi^{(1)}_+ \right), \bar{\xi}\Psi^{(1)}_- \bigg) + 2L_{2}^{(1,0)}\bigg( L_{2}^{(1,0)}\left( \bar{\xi}\Psi^{(1)}_+, \Psi^{(1)}_+ \right), \Psi^{(1)}_- \bigg) \right\rangle_{S, L} \\
&+ \left\langle \bar{\xi}\bar{\mathcal{X}}e^{i}_-, -4L_{2}^{(1,0)}\bigg( L_{2}^{(1,0)}\left( \Psi^{(1)}_-, \Psi^{(1)}_+ \right), \bar{\xi}\Psi^{(1)}_+ \bigg) + 2L_{2}^{(1,0)}\bigg( L_{2}^{(1,0)}\left( \bar{\xi}\Psi^{(1)}_-, \Psi^{(1)}_+ \right), \Psi^{(1)}_+ \bigg) \right\rangle_{S, L} \\
&+ \left\langle \bar{\xi}\bar{\mathcal{X}}e^{i}_-, 2L_{2}^{(1,0)}\bigg( L_{2}^{(1,0)}\left( \bar{\xi}\Psi^{(1)}_+, \Psi^{(1)}_- \right), \Psi^{(1)}_+ \bigg) \right\rangle_{S, L}. \\
\end{split}
\end{equation}
It is then obvious to see
\begin{equation}
\mathcal{O}_{i}^{(3)} = \mathcal{O}_{i,A}^{(3)} + \mathcal{O}_{i,B}^{(3)} = 0,
\end{equation}
which establishes the vanishing of the obstruction.

\bibliography{main}
\end{document}